\documentclass[fleqn,usenatbib,letters]{mnras}

\usepackage[dvipsnames]{xcolor}

\usepackage{newtxtext,newtxmath}

\usepackage[T1]{fontenc}

\def\grs{GRS\,1915+105}
\def\micron{$\mu$m}
\def\spx{{\em SPHEREx}}
\def\jwst{{\em JWST}}
\def\xrism{{\em XRISM}}
\def\nustar{{\em NuSTAR}}
\def\chandra{{\em Chandra}}
\def\xmm{{\em XMM-Newton}}

\def\lmir{$L_{\rm MIR}$}
\def\lx{$L_{\rm X}$}
\def\nh{$N_{\rm H}$}
\def\lintx{$L_{\rm X}^{\rm intrinsic}$}
\def\lintnustar{$L_{\rm 3-78}^{\rm int}$}
\def\llimnustar{$L_{\rm 3-78}^{\rm obs,lim}$}
\def\sigmat{$\sigma_{\rm T}$}
\def\gtsim{\mathrel{\hbox{\rlap{\hbox{\lower3pt\hbox{$\sim$}}}\hbox{$>$}}}}
\def\ltsim{\mathrel{\hbox{\rlap{\hbox{\lower3pt\hbox{$\sim$}}}\hbox{$<$}}}}
\def\Msun{M$_{\odot}$}
\def\supp{$\mathcal{S}(N_{\rm H})$}

\usepackage{graphicx}	
\usepackage{amsmath}	

\title[GRS\,1915+105 sleeping, not just veiled]{A deep intrinsic decline of accretion activity in GRS\,1915+105 during 2025
}
\author[Gandhi and Boorman]{
Poshak Gandhi$^{1}$\thanks{E-mail: poshak.gandhi@soton.ac.uk}
and Peter G. Boorman$^{2,1}$
\\
$^{1}$School of Physics \& Astronomy, University of Southampton, Southampton SO17 1BJ, UK\\
$^{2}$Max Planck Institute for Extraterrestrial Physics, Giessenbachstrasse, 85741 Garching, Germany
}

\date{Revised 2026 Nov 13; Received in original form 2026 Jul 06}

\pubyear{\the\year{}}

\begin{document}
\label{firstpage}
\pagerange{\pageref{firstpage}--\pageref{lastpage}}
\maketitle

\begin{abstract}
The canonical microquasar \grs\ has undergone unprecedented multiwavelength changes since 2018. Hard-X-ray \emph{NuSTAR} observations in November 2025 failed to detect it, reaching flux limits several orders of magnitude below historical X-ray levels. The non-detection could reflect enhanced obscuration, a sharp decline in accretion power, or a combination of the two. Hiding an intrinsically luminous source requires effectively complete coverage by Compton-thick material to limit residual direct or scattered components to $\lesssim$\,10$^{-4}$ of the nuclear power. Free--free absorption by an ionised cocoon could also explain the lack of a self-absorbed compact jet in deep radio imaging in mid-2026. However, such a cocoon would reradiate the absorbed power, contrary to faint infrared emission observed by \emph{SPHEREx} from September 2025 onwards, and consistent with the donor star alone. The source's position in the infrared--X-ray plane also now matches the locus of quiescent black-hole binaries. Together, the X-ray, infrared, and radio constraints exclude any quasi-static, Compton-thick cocoon capable of hiding an otherwise luminous source, and favour an intrinsic quenching in 2025. We propose that strong outflows associated with repeated multiwavelength flaring during 2023--2025 depleted the inner accretion flow, reducing its power by at least two orders of magnitude (rising to $\gtrsim$\,4 dex if the nucleus is now unobscured), and indicating a likely return of \grs\ to quiescence after 33 years, notwithstanding any future rebrightening episodes.  With disc veiling now minimal, the donor star, binary orbit, and residual accretion can be probed directly for the first time.
\\
\end{abstract}




\section{Introduction}

The microquasar \grs\ is the first Galactic source in which apparent superluminal motion was identified, having been active for more than 30 years since its discovery by the {\em Granat} mission in 1992 \citep{castrotirado92, Mirabel1994}. Its bright, long-lived activity and spectacular multiwavelength variability characteristics make it unique amongst Galactic black-hole X-ray binaries \citep{belloni00, Motta2021}. 

In early 2018, the source dimmed dramatically in X-rays and thereafter entered a prolonged low-flux, heavily obscured state \citep{negoro18,miller20, neilsen20, balakrishnan21}. This was punctuated by spectacular multiwavelength flaring and outflows peaking around 2023--2025, during which the source flux soared to peak historical levels in the infrared and radio \citep{sanchezsierras23, Gandhi2025,motta27}. Furthermore, during 2024--2025, the source exhibited some remarkable features in the radio, including continued flaring and significant swings in jet position angle \citep{trushkin25,Rodriguez2025, jiang2026, yan26}. The cause of these unprecedented changes remains unclear, with Lense--Thirring precession, radiation-driven inner disc warps and periastron interactions with the tertiary stellar component of a hierarchical triple having been proposed \citep{Rodriguez2025, miller2025, jiang2026, deregt2026}. 

Following the aforementioned burst of activity, the source has continued to decline in X-rays according to the MAXI all-sky monitor \citep{maxi}. Very recent monitoring has reported non-detections of the source during June 2026 in deep radio imaging \citep{motta26}, and also in soft X-rays \citep{marino26}. These non-detections could arise if either the circumnuclear obscuring column has risen further, or if the source has begun to decline in accretion activity intrinsically. The two interpretations remain degenerate on the basis of these data alone, so the present level of core activity in the source remains uncertain. Following these reports, long-term radio monitoring data was reported by \citet{motta27}, including detections at its faintest levels so far with MeerKAT at two frequencies in July 2026, with a spectral index consistent with optically thin synchrotron emission; the contribution of any self-absorbed compact jet component is limited to be weaker.

Here, we report recent X-ray and infrared observations pushing the multiwavelength detectability constraints to their limits and helping to break this degeneracy. We collate the deepest current limits and detections on the broadband spectral energy distribution (SED), and place these in the context of the recent evolution of the source. We end with a discussion of possible new opportunities afforded in this new phase. A source distance of $d$\,=\,9.4\,kpc and a black-hole mass $M_{\rm BH}$\,=\,11.2\,M$_{\odot}$ \citep{Reid2023} are assumed throughout. Other physical parameters are replicated from Table\,1 of \citet{Gandhi2025} where necessary.

\section{Data}
\label{sec:data}

\subsection{\nustar\ X-ray observations}
\label{sec:nustar}
\grs\ was observed with the {\em NuSTAR} telescope \citep{nustar} on MJD\,60996 (2025 Nov 17 05:05 UTC) for a cleaned exposure time of approximately 32\,ks on both FPMA and FPMB (ObsID 91101345002). This is bracketed in time between two \spx\ epochs described in the following section. The data were reduced using the {\tt NuSTARDAS} v2.1.2 pipeline distributed within {\tt HEASoft v6.30.1}. Calibrated and screened event files for FPMA and FPMB were produced with {\tt nupipeline} and {\tt CALDB v20250616}. Source/background spectra, light curves, and images were extracted with {\tt nuproducts} following standard {\em NuSTAR} analysis procedures \citep{madsen15}. Since no clear source was visible in the {\em NuSTAR} images, products were extracted from circular regions assuming nominal J2000 coordinates of the source (see \S\,\ref{sec:spx}) with an `on' region extraction radius of 49\,arcsec and `off' region radii as large as possible on the same detector as the source, for each FPM. The mean observed count rates in the source region (SR) and background region (BR) are $C_{\rm SR}$\,=\,6.7\,($\pm$\,0.3)\,$\times$\,10$^{-3}$\,ct\,s$^{-1}$ and $C'_{\rm BR}$\,=\,7.2\,($\pm$\,0.1)\,$\times$\,10$^{-3}$\,ct\,s$^{-1}$ (detector averages with  1$\sigma$ uncertainties), respectively, where the prime indicates that the rate has been scaled to account for the differing aperture sizes of the two regions. The count rates are entirely dominated by background.

Quantifying the flux upper limit relies on an accurate characterisation of the background extracted at the expected location of the source. We fit the unbinned source\,$+$\,background and background spectra for both FPMs simultaneously across 3--78\,keV using the Poisson likelihood {\tt cstat} \citep{Cash79} and the Bayesian X-ray Analysis software package ({\tt BXA v2.9}; \citealt{Buchner14,Buchner21}). {\tt BXA v2.9} connects the Python wrapper of the nested sampling algorithm {\tt MultiNest v3.10} \citep{Feroz09,Feroz19} to the Python wrapper of the X-ray spectral fitting software {\tt Xspec v12.12.1} \citep{Arnaud96,Gordon21}. Our use of nested sampling ensures an accurate quantification of the maximum flux the source can exhibit for a variety of different model shapes whilst still being sub-dominant to the background (see e.g., \citealt{Buchner23a,Buchner23b}). To ensure accurate reconstruction of parameter posteriors, we use a sampling efficiency of 0.1 with 1000 live points \citep{Dittmann24}. The PCA background models of \citet{Simmonds18} were then fit to derive background spectral models for both FPMs. To include some realistic uncertainty in the background estimation, we include a log-uniform prior to describe the normalisation of each background model between 0.1\,--\,10$\times$ the expected value. The source spectrum was tied between both FPMs and assumed to follow a power law with a log-uniform prior between 10$^{-10}$\,--\,1 and uniform prior between 1\,--\,1.5 to describe the normalisation and photon index $\Gamma$, respectively. The range of possible photon indices was chosen to be representative of the effective shape of an obscured power law. The model fit thus relied on four free parameters, and flux upper limits were measured as the 99.7th percentile of their respective one-dimensional marginalised posterior distributions.

The upper limit on the 3--78\,keV flux derived is $F_{3-78}^{\rm lim}$\,$<$\,1.3\,$\times$\,10$^{-13}$\,erg\,s$^{-1}$\,cm$^{-2}$ which corresponds to a luminosity upper limit of \llimnustar\,$<$\,1.4\,$\times$\,10$^{33}$\,erg\,s$^{-1}$. This is an estimator of the observed limit without correction for any intrinsic absorption, and is about 10$^{4-6}\times$ deeper than the range spanned by historical X-ray detections of the source, including some taken during the 2018--2025 X-ray-obscured phase (see compilation in Fig.\,9 of \citealt{Gandhi2026}). We will consider detailed spectral modelling constraints in \S\,\ref{sec:discussion_cocoon}, and simply note here that this basic model flux limit is conservative (i.e., higher) with respect to typical unabsorbed models, including power-law shapes characterised by softer $\Gamma$, or a multicolour disc blackbody with effective temperature $T$\,$\sim$\,1\,keV, say.

\subsection{SPHEREx infrared spectrophotometry}
\label{sec:spx}
The `{\em Spectro-Photometer for the History of the Universe, Epoch of Reionization, and Ices Explorer}' (hereafter, \spx; \citealt{spxmission}) is a 20-cm effective-aperture mission, conducting the first all-sky spectral imaging mapping survey in the infrared between 0.75--5.0\,\micron. It has a low spectral resolution but its all-sky coverage and high sensitivity --- approximately similar to the photometric sensitivity of the {\em WISE} mission \citep{wright10} at the same wavelengths --- enable serendipitous discoveries and monitoring across the entire sky. 

\begin{figure}
\centering
\includegraphics[width=1.0\linewidth]{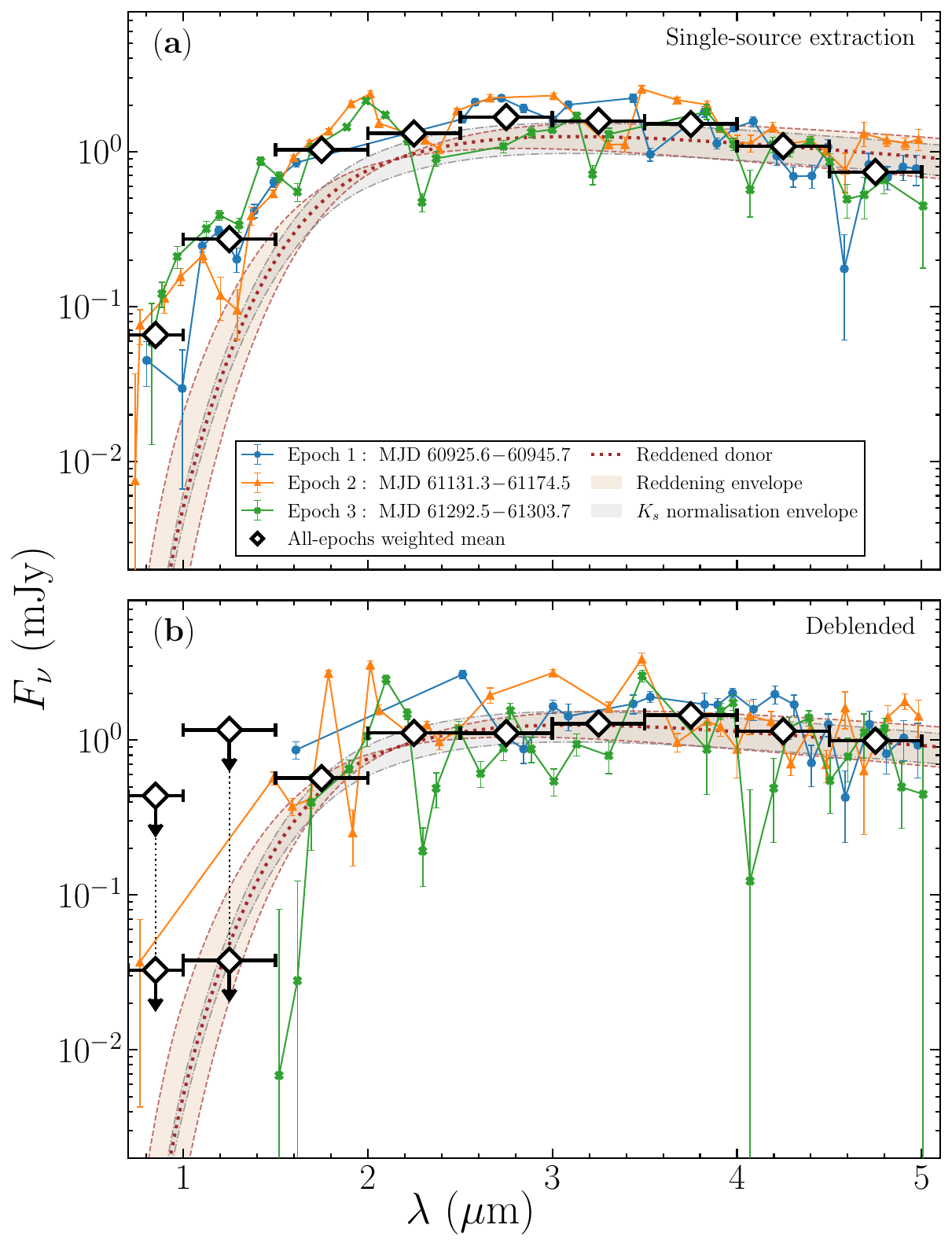}
\caption{(a) Average observed single-source SEDs (across the \spx\ bands) of GRS\,1915+105, split into the three observing epochs (blue, orange and green points, respectively). (b) Deblended SEDs following the simultaneous source fit described in the text. Uncertainties plotted are representative of 68\,\% confidence. Negative fluxes are preserved in the analysis but not plotted on the logarithmic scale. The large black-outlined diamond symbols represent weighted flux-density averages across all three epochs and grouped in coarse wavelength bins. Grouped bins with negative mean flux densities are treated as upper limits illustrated as arrows denoting both the standard error (lower arrows) and standard deviation (upper arrows) of relevant flux bins, connected by a vertical dotted black line. In both panels, the dotted brown line is a $T$\,=\,4,000\,K blackbody (K-giant donor) normalised at $K$\,$\approx$\,14.75\,$\pm$\,0.25 mag,  $F_\nu$\,$\approx$\,0.9\,$\pm$\,0.2\,mJy \citep{Greiner2001}, and reddened with the tabulated extinction law from \citet{Gandhi2025}, with $A_{\rm V}$\,=\,19.6\,$\pm$\,1.7\,mag. The red and blue envelopes plotted around the donor spectrum denote the effect of these quoted reddening and normalisation uncertainties, respectively. 
\label{fig:spx}}
\end{figure}

We utilised the \spx\ Quick Release 2 (QR2) and
Quick Release 3 (QR3) Level-2 spectral image products. QR2 became public in October 2025 and included reprocessed earlier observations, whereas QR3 began release in September 2026 and incorporates observations from the third and fourth all-sky surveys, together with improved point-spread-function and photometric calibrations. These releases are described in the associated Explanatory Supplement\footnote{\spx\ Explanatory Supplement, v2.0 (SSDC-DP-017): \url{https://irsa.ipac.caltech.edu/data/SPHEREx/docs/SPHEREx_Expsupp_QR_v2.0.pdf}}. Differences in calibration between QR2 and QR3 warrant caution when interpreting small inter-epoch flux variations; but our key conclusions below rest on a stronger and systematic infrared faintness of the source, and remain robust to current calibration uncertainties.

To begin with, all available single-source spectrophotometric measurements of the source were extracted using the Spectrophotometry tool provided by the mission team.\footnote{https://irsa.ipac.caltech.edu/applications/spherex/tool-spectrophotometry} We adopted the radio-VLBI position of \grs, propagated to the mean \spx\ observing epoch using the variance-weighted proper motion reported in \citet{Reid2023}. This gives an ICRS/J2000 position of ($19^{\rm h}15^{\rm m}11.54344^{\rm s}$, $+10^\circ56'44\farcs 5863$). Pipeline processing encodes pixel-level quality information in a flag map for each detector array, identifying pixels that need to be excluded in photometric flux summation. These flags include a wide range of artefacts such as hot, cold and nonlinear pixels, those impacted by overflow errors, persistence effects, and more. We retained only flags equal to zero or to the nominal source-mask value $2^{21}$ as defined in the Explanatory Supplement, thereby excluding measurements affected by propagated pixel-quality flags or fitting failures. Experimenting with slightly different flag selection did not change the final extraction significantly. The spectrophotometry tool utilises the {\tt Tractor} algorithm \citep{tractor} to conduct PSF-convolved forced photometry in individual channel images at the source location, which is then collated across all available images to construct an aggregated source spectrum. Background was measured in a large annulus around each source position. 

We show the single-source-extraction SEDs in the top panel of Fig.\,\ref{fig:spx}, split according to observing epoch. \spx\ builds up spectrophotometric observations of a source over time, with observing epochs typically lasting a few weeks during which effective wavelengths across its full bandpass are cycled through as a function of detector position. For \grs, three \spx\ observing epochs are available, spanning MJD 60925--60945 (2025 September 7--27), MJD 61131--61174 (2026 April 1--May 14), and MJD 61292--61312 (2026 September 9--29), respectively. The fluxes between the epochs are broadly consistent with each other. There are subtle differences between individual filters, but these lack any obvious pattern. At this early stage of the mission and pipeline processing, subtle variations also need to be treated with some caution. 

An absorbed blackbody is also  overplotted to represent the donor, illustrative of a K5\,III donor \citep{Steeghs2013}, normalised to an absorbed $K$-band flux density of $\approx$\,0.9\,$\pm$\,0.2\,mJy \citep{Greiner2001}. An extinction of $A_{\rm V}$\,=\,19.6\,$\pm$\,1.7\,mag is adopted, with the reddening law tabulated in \citet{Gandhi2025}.\footnote{This was derived by combining a selective-extinction coefficient $R_{\rm V}$\,=\,3.1 with the published optical--to--near-IR and near--to--mid-IR reddening laws of \citet{fitzpatrick09} and \citet{Chiar2006}, respectively.} The donor can satisfactorily account for the observed fluxes at the longer wavelength end, especially when considering systematic uncertainties in extinction and donor normalisation. But there is an apparent excess in the data above donor-star expectations at  $\lambda$\,$\lesssim$\,3\,\micron. If real, this could be indicative of residual accretion flux whose contribution rises toward the blue. However, other possible contributing biases ought first to be tested.

The field around \grs\ is crowded in the infrared, and this is the primary caveat in analysing faint sources with \spx, because of its relatively coarse pixel scale of 6\farcs 15. Deblending errors are known to be a key limitation in current processing \citep{bahk26}. This can be refined through {\em multi-source} spectrophotometry. In this mode, the coordinates for multiple sources are passed to the {\tt Tractor} workflow, which conducts simultaneous multi-source forced photometric fitting at all specified locations, allowing a better deblending of individual objects in the process. Candidate contaminating sources within 60$''$ of the aforementioned single-source position of \grs\ were assembled from the UKIDSS Galactic Plane Survey, {\em Spitzer}/GLIMPSE and 2MASS catalogues \citep{Lucas2008,Benjamin2003,Churchwell2009,Skrutskie2006}. UKIDSS provides the highest angular resolution, while GLIMPSE and 2MASS supplied longer-wavelength and bright-source constraints. Catalogue detections were cross-matched and de-duplicated. {\em Gaia} DR3 counterparts \citep{gaiadr3} with reliable astrometry were propagated from their J2016.0 reference epoch to the mean \spx\ epoch, MJD\,61125.68 (2026 March 26). The \spx\ archive currently permits a maximum of 20 sources in its simultaneous fit. We initially included every validated source whose centroid lay within 6\farcs15 of \grs. A total of 8 UKIDSS detections lie within 6\farcs15 (i.e., within a distance corresponding to one \spx\ detector pixel) of \grs\ down to a mag $K$\,=\,18.3. The two brightest neighbours most relevant for deblending are UKIDSS sources 438932193009 (WFCAM mag $K$\,=\,13.3) and 438932193063 ($K$\,=\,15.2) at separations of 4\farcs6 and 3\farcs2, respectively, with all remaining being $\ge$\,1\,mag fainter still. Three of the fainter sources resulted in strongly covariant solutions with \grs\ in the deblending; following experimentation, we found that excluding these yielded reliable and stable photometric deblending solutions. At larger separations, candidate contaminants were ranked according to their near-/mid-infrared brightness and likely contribution to extended PSF wings relevant for profile-fitting photometry. A total of 16 neighbours were thus fit simultaneously with \grs.

The resultant deblended \spx\ SED of \grs\ is shown in the bottom panel of Fig.\,\ref{fig:spx}. Despite larger uncertainties than the single-source extraction, there is now much better agreement between the data and companion-star spectrum, implying nominal deblending success. Negative fluxes are preserved for relevant analysis. We find that the short-wavelength deblended mean fluxes end up being negative and formally consistent with zero; this could be attributable to deblending over-corrections especially at the shorter near-IR end, where the contribution of less-reddened foreground stars is likely to be higher than at longer wavelengths.

There is no evidence for significant residual excess above the reddened donor-star model, with a maximum significance of only $1.14\sigma$. The mean measured excess-to-donor ratio across the wavelength range is $\simeq$\,0.19, while the least constraining bin permits a conservative additional component of up to $1.71\,F_{\nu,\,{\rm donor}}$. Thus, there is no residual evidence of underlying significant accretion signatures, to within the limits of the present data. Irrespective of this, deblending uncertainties do not impact our final inferences below, and for the purposes of the analysis below, we utilise the deblended SED across all three epochs and grouped in coarse bins (large diamonds in the figure), with non-detections treated as conservative standard-deviation upper limits.

\section{Discussion}
\subsection{Broadband Spectral Energy Distribution}
\label{sec:discussion_sed}
\begin{figure*}
\centering
\includegraphics[width=0.9\textwidth]{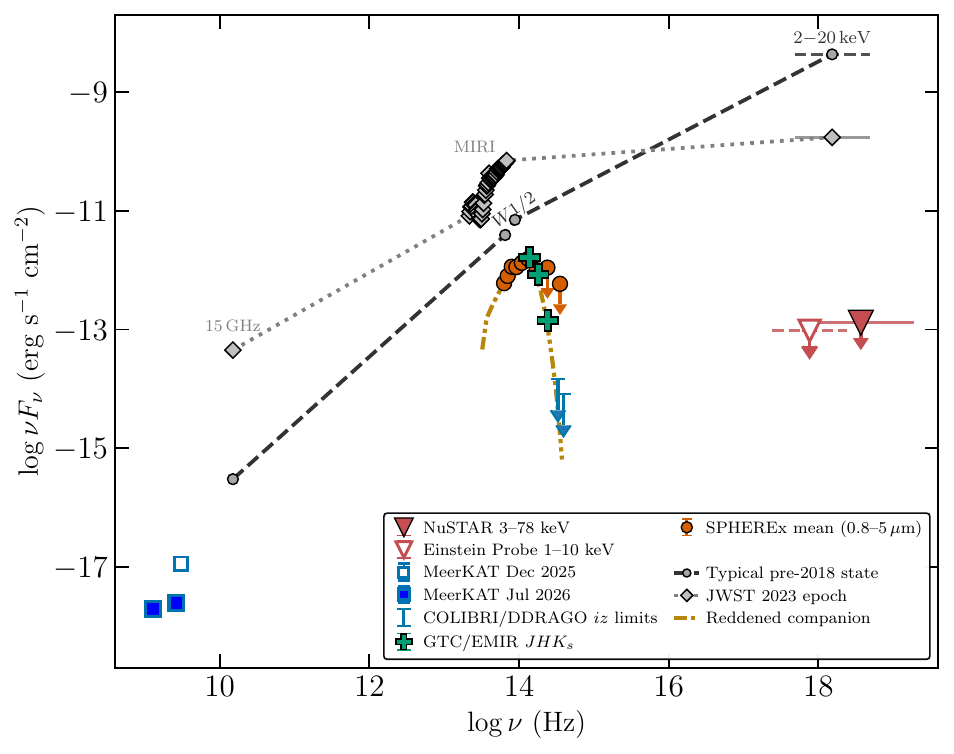}
\vspace{-0.3cm}
\caption{Broadband SED of \grs\ in $\nu F_\nu$ units. The recent deep detections and limits shown in colour sample the period from late 2025 through to mid-2026; from right to left, these are \nustar\ (hard X-rays), EP (soft X-rays), COLIBRI (optical), GTC (near-IR), \spx\ (near and mid-IR), and MeerKAT (radio), as described in the text. Two comparison SEDs are shown: (1) a typical SED from September 2017, before the onset of the X-ray obscured state, is shown as the dashed line connecting grey circles; (2) the SED during strong multiwavelength flaring coordinated around {\em JWST}/MIRI in mid-2023 is represented by the dotted line connecting grey diamonds. Flux densities for both comparison SEDs are from \citet{Gandhi2025} and references therein. The mean expected companion-star SED from Fig.\,\ref{fig:spx} is depicted as the dot-dashed curve. 
\label{fig:sed}}
\end{figure*}

Fig.\,\ref{fig:sed} shows the broadband spectral energy distribution (SED) of the source based on the data in \S\,\ref{sec:data}, combined with five additional sets of reported multiwavelength detections and limits that have recently become available: (1) optical limits from COLIBRI/DDRAGO of $i$\,$>$\,23.12 and $z_y$\,$>$\,22.30 (AB, 3$\sigma$) from MJD 61227 \citep{ducoin26}; (2) near-infrared detections of $J$\,=\,18.58\,$\pm$\,0.07, $H$\,=\,15.85\,$\pm$\,0.07 and $K_{\rm s}$\,=\,14.37\,$\pm$\,0.06 (Vega) from MJD 61226 \citep{cotizelati26}; (3) X-ray non-detections with Einstein Probe (EP) from MJD 61218--61253 \citep{motta27, marino26}; (4) deep MeerKAT detections at $150\pm34\,\mu$Jy (1.28\,GHz) and $93\pm11\,\mu$Jy (2.62\,GHz) in 2026 July (MJD 61228--61234) \citep{motta27}; and (5) a deep MeerKAT S-band detection of $368\pm37\,\mu$Jy at 2.62\,GHz on MJD 61011 \citep{motta27}, paired by the authors with the \nustar\ non-detection for interpretation. Standard photometric zeropoints were utilised to convert magnitudes to fluxes, where necessary. The EP X-ray flux limit includes a small conversion factor to the same photon index as for \nustar.

Two comparison SEDs are also overplotted. The first is an SED from data gathered around MJD\,58000 (2017 September 04), pre-dating the 2018--2025 X-ray obscured state, and approximately representative of the historical source multiwavelength appearance across decades. The second is an SED from around MJD 60101, corresponding to an epoch of multiwavelength coordination around a {\em JWST} observation reported in \citet{Gandhi2025}, when the source was caught during a period of unprecedented infrared flaring. Assuming reasonable flux extrapolations between comparison frequencies where relevant (e.g., in the radio), the source is substantively fainter now than in the past. This is true across the electromagnetic spectrum. It is also noteworthy that the GTC and \spx\ infrared detections all independently and closely match the expected companion-star flux levels.

The non-simultaneity of these collated recent multiwavelength datasets should be kept in mind for some of the detailed interpretations and extended geometries discussed below. Furthermore, the radio detections could potentially be sampling vastly different spatial scales to the other bands; as discussed by \citet{motta27}, significant contributions of extended past ejecta cannot be excluded at the spatial resolution limit of MeerKAT. However, the recent fluxes and detection limits are all at the faintest level in the history of \grs, and relevant response times --- including light-crossing, dynamical and recombination timescales --- are typically of order $\mathcal{O}({\rm seconds\,\sim\,days})$ for much of the parameter space discussed (dynamical timescales approach $\sim$\,1\,yr only for $R$\,$\gtrsim$\,10$^{14}$\,cm). These timescales are substantially shorter than the period from September 2025 to mid-2026 across which observations were gathered. Furthermore, there is broad agreement between datasets separated by many months, including the GTC and \spx\ infrared detections, as well as the \nustar\ and EP X-ray non-detections (albeit with only slight overlap in energy coverage). Further tests are discussed in later sections, but we do not consider time variability to be a factor limiting our key interpretations.

\begin{figure*}
\centering
\includegraphics[width=0.75\textwidth]{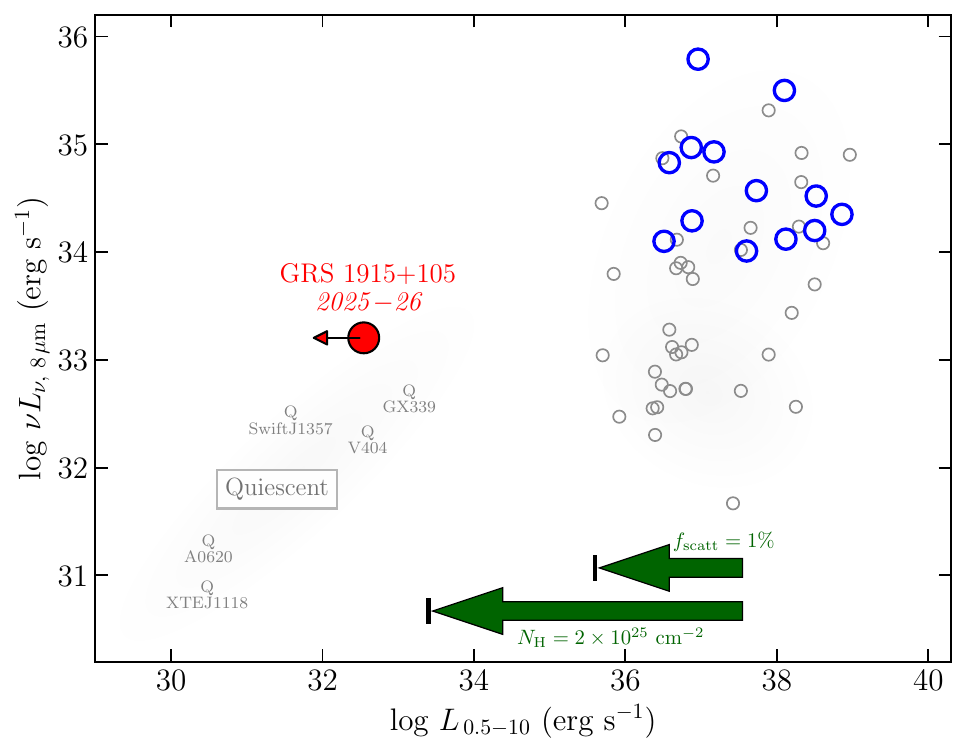}
\caption{The \lmir--\lx\ luminosity plane for X-ray binaries, with \grs\ historical observations in the hard and obscured states in blue, and the 2025--26 \nustar/\spx\ epoch highlighted as the large red circle. The plotted 8\,\micron\ flux density is extrapolated from the empirical 3.5--5\,\micron\ \spx\ mean SED which approximates a power law with $F_\nu$\,$\propto$\,$\nu^{1.75}$. The \spx\ detection is fully consistent with the companion star in this case, and any accretion-related infrared component is even weaker. This is an observed-luminosity diagnostic plot; X-ray luminosities are derived from the observed fluxes, corrected for interstellar absorption only. The thick green arrows denote X-ray flux decrements by factors of 100 and $\approx$\,10$^4$ (upper and lower arrows, respectively, with the latter originating in a Compton-thick column \nh\,=\,2\,$\times$\,10$^{25}$\,cm$^{-2}$ for the {\tt BNsphere} obscurer model) relative to the median flux level observed historically. \grs\ now lies in the regime consistent with quiescent-state sources. Adapted from \citet{Gandhi2026}, with the remaining sources in that compilation shown as light-grey circles (hard/soft) and with the letter `Q' and shortened source-name identifiers for quiescent sources. 
\label{fig:lmirlx}}
\end{figure*}

Fig.\,\ref{fig:lmirlx} displays the current location of the source on the mid-infrared--X-ray plane of black-hole X-ray binaries from \citet{Gandhi2026}. The plotted X-ray limit is $L_{0.5-10}$\,$\le$\,3.5\,$\times$\,10$^{32}$\,erg\,s$^{-1}$, derived from the observed \nustar\ flux limit under our power-law ansatz model (with conservative $\Gamma$\,=\,1.5; \S\,\ref{sec:nustar}) corrected only for interstellar gas column density of \nh\,=\,1.4\,$\times$\,10$^{22}$\,cm$^{-2}$ (\citealt{h4pi}); adopting instead the $\approx$\,4\,$\times$\,10$^{22}$\,cm$^{-2}$ implied by our $A_{\rm V}$ \citep{guver09} raises this by only $\approx$\,0.1\,dex, since the limit is set above 3\,keV. The stark decline in the source with respect to its historical behaviour and also with respect to other sources in typical X-ray `hard' and `soft' accretion states is immediately apparent, irrespective of residual deblending uncertainties. Since the infrared detections are consistent with companion-star emission, a conservative interpretation would treat the plotted infrared flux level as an upper limit on any ongoing accretion activity.

\subsection{Enhanced obscuration?}
\label{sec:discussion_cocoon}

We turn next to the question of whether the broadband changes can be attributed solely to an enhanced obscuring screen. The {\em NuSTAR} non-detection is strongly constraining in this regard. Near-Compton-thick levels of obscuration were inferred to be veiling the source during the early phases of the X-ray obscured state \citep{miller20, Motta2021, balakrishnan21}. The {\em hard} X-ray focusing capability of {\em NuSTAR} above 10\,keV makes it exquisitely sensitive to any flux leaking through such obscuring columns, in the regime where Compton scattering begins to dominate over photoelectric absorption \citep{rybickilightman}.

The total X-ray optical depth at energy $E$ is a combination of absorption and scattering of photons out of the line-of-sight, with respective optical depths written as
\begin{eqnarray}
    \tau_{\rm X}(E)&=&\tau_{\rm abs}(E)+\tau_{\rm scatt}\nonumber\\
    &\simeq&(1-x)\tau_{\rm PE}(E)+N_{\rm H}\sigma_{\rm T}.
\end{eqnarray}
$x$ represents the ionisation fraction of the gas. If a non-negligible fraction of the screening gas is neutral or partly ionised, photoelectric (PE) absorption is the primary process depleting photons from the line-of-sight, especially below $E$\,$\sim$\,10\,keV. In the harder X-ray band $E$\,$\gtrsim$\,10\,keV, Compton scattering dominates $\tau_{\rm X}$, and its optical depth approximates that of Thomson scattering, $\tau_{\rm scatt}$\,$\approx$\,$\tau_{\rm T}$\,$\approx$\,\nh\,\sigmat. Formally, $\tau_{\rm scatt}$ is also a function of energy, but the relevant Klein--Nishina scattering corrections are small in the \nustar\ regime and are thus ignored here, rendering $\tau_{\rm scatt}$ approximately independent of $E$. 

$\tau_{\rm X}$ rises rapidly at all X-ray energies above \nh\,$\approx$\,\sigmat$^{-1}$\,=\,1.5\,$\times$\,10$^{24}$\,cm$^{-2}$, and  \nustar\ is particularly sensitive to photons leaking through such Compton-thick obscuring columns. Two of the most deeply obscured active galactic nuclei (AGN) where \nh\ has been reliably estimated with \nustar\ --- Circinus Galaxy and NGC\,1068 --- show \nh\,$\gtsim$\,10$^{25}$\,cm$^{-2}$ \citep{Arevalo14,Bauer2015,Boorman2024}. At this column, scattering would reduce the directly-transmitted photon intensity to a fraction 
\begin{equation}
    f_{\rm transm}=e^{-10^{25}\times\,\sigma_{\rm T}}=e^{-6.7}\approx10^{-3}
\end{equation}

\noindent
relative to the emitted intensity. The apparent flux reductions that we are now seeing in \grs\ are even more extreme (corresponding to $f_{\rm transm}$\,$\sim$\,10$^{-4}$--10$^{-6}$; cf. \S\,\ref{sec:discussion_sed}, Fig.\,\ref{fig:lmirlx}) and thus require even higher total optical depths, in the range $\tau_{\rm X}$\,$\sim$\,9--14.

It is possible to place additional constraints on any obscurer, based on geometrical arguments. Even when direct sightlines to the core are Compton-thick, indirect scattering {\em into} the line-of-sight is expected to produce detectable flux across a wide X-ray bandpass. Pointed high-sensitivity {\em XRISM}/Resolve microcalorimeter observations of \grs\ as late as October 2024 revealed a distinct scattered emission component \citep{miller2025}. Suppressing this component would require a further increase in the obscurer's covering factor, or additional attenuation of the scattering region. The mature literature on nearby heavily obscured AGN again provides a useful empirical comparison, with X-ray scattering fractions $f_{\rm scatt}$ generally ranging from below 1 per cent to a few per cent \citep[e.g.,][]{Bauer2015,Boorman2025,Gupta2021,Ricci17_bassV}. Such scattering fractions have also been successfully accommodated on a population level within AGN synthesis models of the cosmic X-ray background radiation \citep{Comastri1995, gandhifabian03, gandhi2007, Gilli2007, ueda14, ananna19}. If an obscurer with a similar scattering geometry were to exist around \grs, we would then expect the \nustar\ observation to show at least some detectable flux, even if Compton-thick. The fact that we do not see this implies that any obscuring cocoon would need to have exceptionally low $f_{\rm scatt}$, equivalent to a 4$\pi$-covering geometry that is also Compton-thick in {\em all} directions, to within 1 part in $\sim$\,10$^{4-6}$, if the intrinsic accretion-power output remains undiminished.

\begin{figure*}
    \centering
    \includegraphics[width=0.46\linewidth]{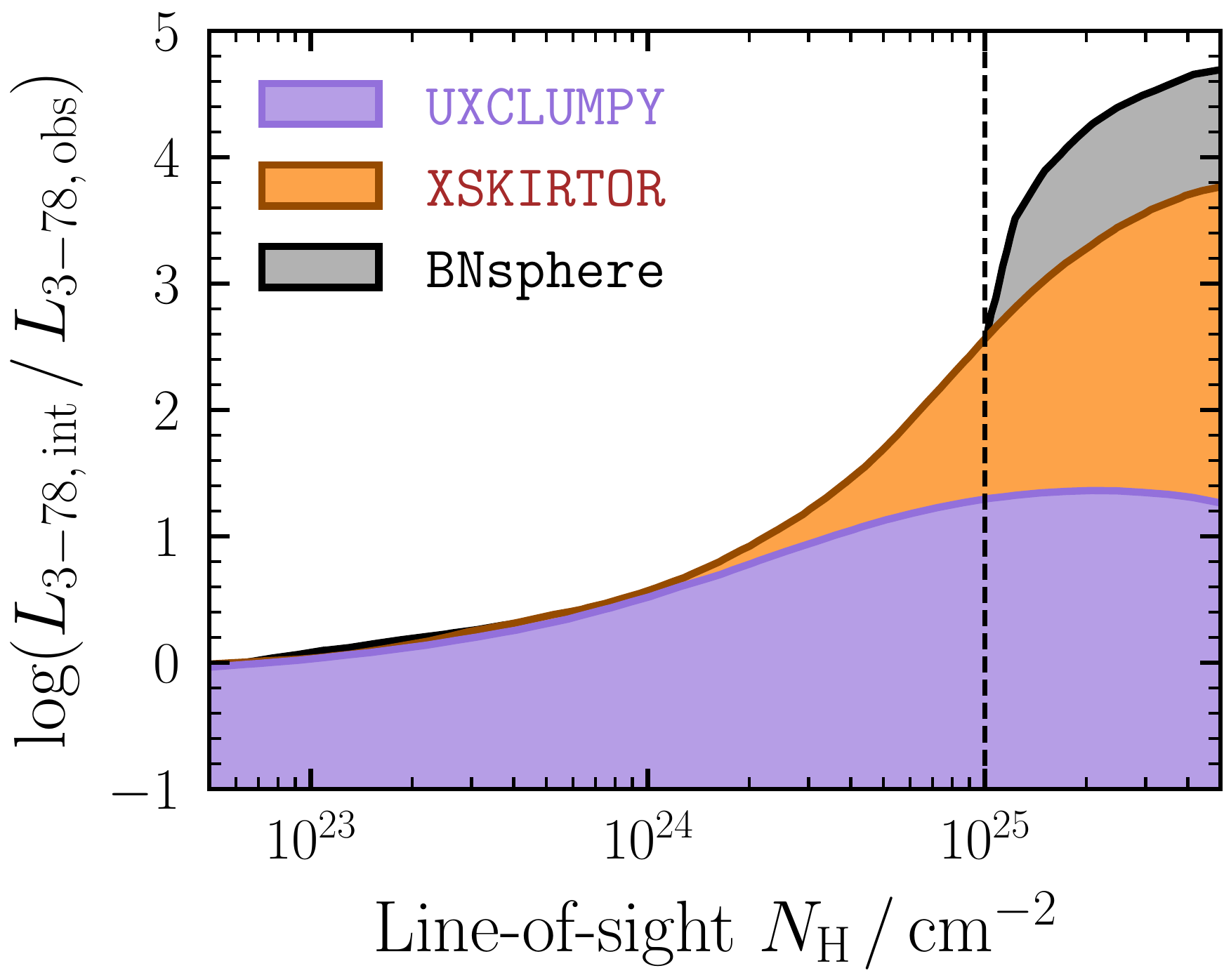}
    \hfill
    \includegraphics[width=0.48\linewidth]{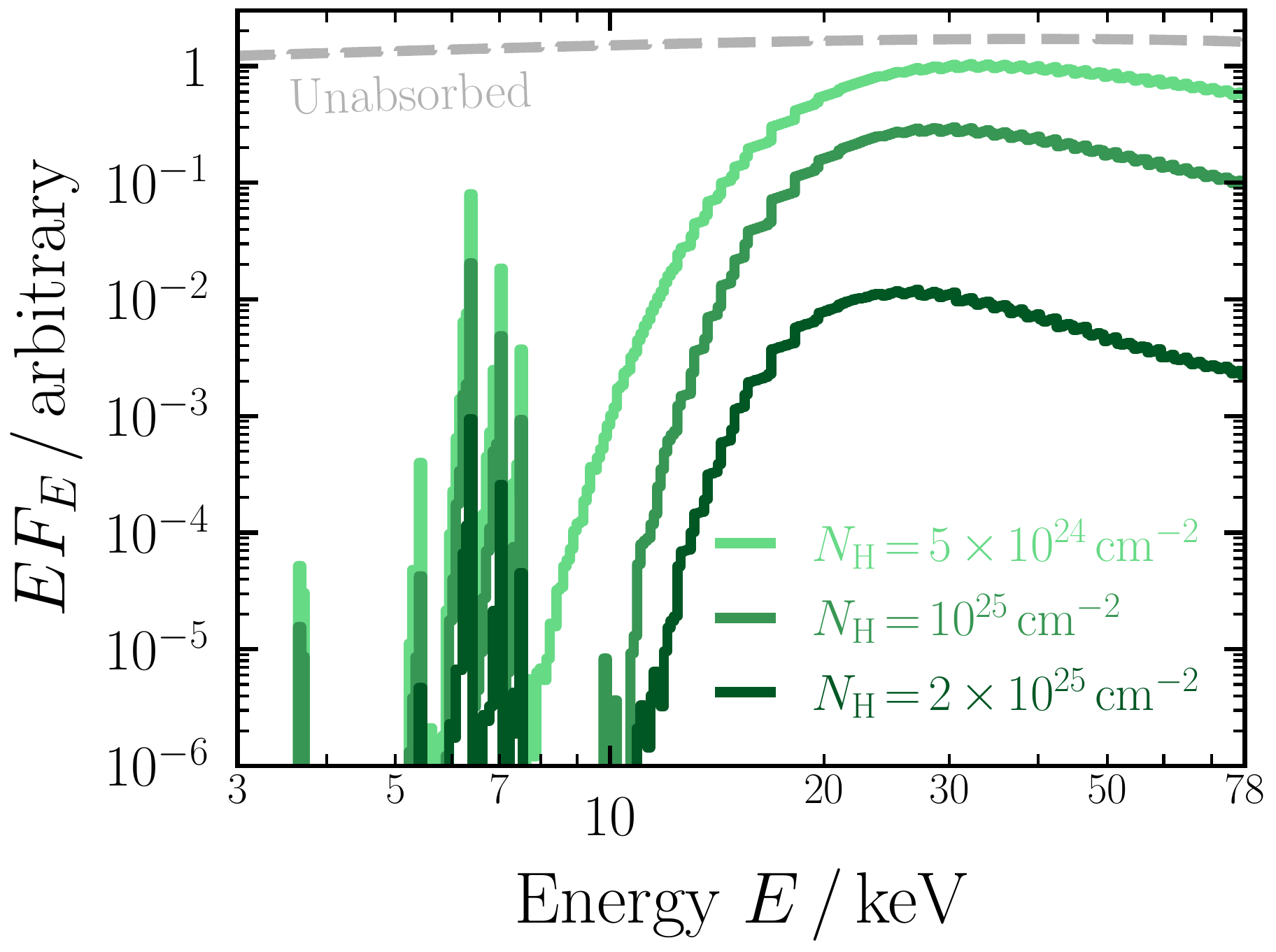}
    \caption{{\em (a) Left panel}: X-ray obscurer models and the impact of geometry and column density on the resultant suppression (\supp) of X-ray emission in the 3\,--\,78\,keV passband. The y-axis shows the flux decrement (intrinsic/observed), as a function of line-of-sight column density on the x-axis. {\em (b) Right panel}: Representative (observed) X-ray model spectra produced by the {\tt BNsphere} model as a function of line-of-sight column density. The dramatic decline in the emergent X-ray spectrum, relative to the driving (unabsorbed) continuum, is apparent as \nh\ rises.}
    \label{fig:xmodels}
\end{figure*}

To assess this hypothesis against realistic point-source obscurer scenarios, we repeated the spectral fits described in Section~\ref{sec:nustar} with a number of different physically self-consistent models previously used to describe AGN X-ray absorbers. Our choices are the clumpy obscuration distribution described by \texttt{UXCLUMPY} \citep{Buchner2019}, the smooth obscuration distribution described by \texttt{XSKIRTOR} \citep{VanderMeulen23} and the spherically-covered smooth obscuration distribution of \texttt{BNsphere} \citep{Brightman11a}. For all spectral fits, we fix the intrinsic photon index to 1.8, the high-energy cut-off to $E_{\rm cut}$\,=\,200\,keV and the viewing angle to maximally edge-on. All other parameters describing the global covering factor of the respective obscurers were left free to vary, as well as a small fraction of escaping flux in the \texttt{UXCLUMPY} and \texttt{XSKIRTOR} model set-ups. We find that only the 4$\pi$ covering fraction model \texttt{BNsphere} is capable of producing a flux suppression \supp\,=\,\lintnustar/\llimnustar 
of $>$\,4 orders of magnitude in 3\,--\,78\,keV, and that this only occurs for a line-of-sight column density $N_{\rm H}$\,$\gtrsim$\,2\,$\times$\,10$^{25}$\,cm$^{-2}$. Fig.\,\ref{fig:xmodels} illustrates these model constraints. 

The large spread between models at $N_{\rm H}$\,$\gtrsim$\,10$^{25}$\,cm$^{-2}$ (Fig.\,\ref{fig:xmodels}) is not a numerical systematic, but a direct consequence of uncertainties in the geometry. Once the line-of-sight column density enters the Compton-thick regime, the transmitted continuum becomes negligible and the observed flux is instead set by the floor of photons Compton-scattered or reflected towards the observer by any illuminated material that remains directly visible. In {\tt UXCLUMPY}, illuminated clouds remain visible through gaps between clumps, so this floor is high and \supp\ saturates at $\approx$\,1.4\,dex below the incident power regardless of the line-of-sight \nh. The {\tt XSKIRTOR} torus is smooth, thus exposing less illuminated surface with a correspondingly lower floor (\supp\,$\approx$\,3.8\,dex at 5\,$\times$\,10$^{25}$\,cm$^{-2}$). {\tt BNsphere} encloses the source completely ($\Omega/4\pi$\,=\,1), leaving no exposed reprocessing surface and thus no floor. For each model, the maximum suppression is attained at the largest covering fraction its parameters allow, so that across the three models the achievable suppression rises monotonically with covering fraction.

For reference, we also note that the estimated X-ray flux limit under \texttt{BNsphere} corresponding to our nominal \nustar\ count-rate limits is $F_{3-78}^{\rm lim}$\,$<$\,1.0\,$\times$\,10$^{-13}$ erg\,s$^{-1}$\,cm$^{-2}$. Furthermore, we also tested the impact of a smaller high-energy cut-off as has been inferred in this source (e.g., \citealt{Athulya2023}): experimenting with $E_{\rm cut}$ ranging from 10\,keV (the lower bound allowed by \texttt{BNsphere}) to $\sim$\,100\,keV, we find that $F_{3-78}$ remains below $\approx$\,7\,$\times$\,10$^{-14}$ erg\,s$^{-1}$\,cm$^{-2}$. These various flux limits are all less conservative than our nominal value from \S\,\ref{sec:nustar}, so we retain our original flux limit for the main discussion and Figs.\,\ref{fig:sed} and \ref{fig:lmirlx}.

There are several classes of AGN proposed to have similarly large covering factors, e.g., Optically Quiescent Quasars \citep{Greenwell2021,Greenwell22,Greenwell24, Gandhi2002} or buried low-scattering-fraction AGN \citep{Ueda2007,Eguchi2009}, but there are no systems with 4\,$\pi$ covering fractions that are confirmed at X-ray energies to have equivalent hydrogen column densities $N_{\rm H}$\,$\gtrsim$\,2\,$\times$\,10$^{25}$\,cm$^{-2}$. Possible candidates include the classes of `buried' AGN \citep{imanishi07} and other Compact Obscured Nuclei \citep{Falstad21} or the recently uncovered subset of Little Red Dots suggested to be engulfed in Compton-thick reservoirs of gas (e.g., \citealt{Naidu25,Rusakov2026}). 

Fully embedded central engines have also been proposed in the stellar-mass regime, though optical and infrared observations are typically more constraining than X-rays for these. OH/IR stars --- extreme AGB stars enshrouded by optically thick circumstellar envelopes --- are optically invisible yet among the brightest mid-infrared objects in the Galaxy, with the stellar luminosity fully reprocessed by dust \citep{habing96}. Stellar-merger remnants such as V1309\,Sco and CK\,Vul similarly cocooned themselves in ejecta following their eruptions, fading optically while sustaining strong infrared and sub-mm emission from their reprocessing shells \citep{tylenda11, kaminski15, tylenda16}. Even the failed-supernova candidate N6946-BH1, in which a $\sim$$25\,M_{\odot}$ supergiant vanished optically, retains a bright mid-infrared counterpart of uncertain origin 14 years after the putative event \citep{adams17, beasor24}. Eruptive dust-ejection events around evolved stars show the same behaviour: Sakurai's Object (V4334\,Sgr) and its analogue WISE J180956.27$-$330500.2 became optically enshrouded following massive envelope ejections, yet correspondingly \emph{brightened} dramatically in the mid-infrared as the reprocessing shells expanded and cooled \citep{duerbeck97, gandhi12}. The absence of any analogous infrared excess in GRS\,1915+105 (\S\,\ref{sec:spx}) stands in 
contrast to all of the above.

We also note that the {\tt BNsphere} model suppresses the entire observable continuum for \nh\,$\gtrsim$\,6\,$\times$\,10$^{25}$\,cm$^{-2}$, and it remains unclear whether this is due to physical obscuration suppression, an artefact of Monte Carlo noise arising from limitations of the simulations used to create the model or a combination of both. Conservatively, our spectral fits to the {\em NuSTAR} non-detection of \grs\ with \texttt{BNsphere} are capable of reliably producing a 3\,--\,78\,keV flux suppression of $\sim$\,6\,$\times$\,10$^{4}$ (cf. Fig.~\ref{fig:xmodels}). But higher suppression factors $\sim$\,$\mathcal{O}$(10$^{6}$) required to reach the median level of historical source fluxes (Fig.\,\ref{fig:lmirlx}) are also expected analytically, independent of Monte-Carlo statistics. As a worked example, consider $N_{\rm H}$\,$\approx$\,2\,$\times$\,10$^{25}$\,cm$^{-2}$ ($\tau_{\rm T}$\,$\approx$\,13), for which the unscattered beam is attenuated by $e^{-\tau}$\,$\sim$\,10$^{-5.6}$, while photons escaping by diffusion undergo $N$\,$\sim$\,$\tau^2$\,$\approx$\,170 scatterings \citep[e.g.][]{rybickilightman}. For scattering off cold electrons, the cumulative Compton recoil gives $1/E_N$\,=\,$1/E_0$\,+\,$N/m_ec^2$, so that photons of $E_0$\,=\,30--78\,keV are degraded to $E_N$\,$\approx$\,$m_ec^2/N$\,$\approx$\,3\,keV during their random walk, at which point photoelectric absorption efficiently absorbs them. Hard X-rays are thus removed from the \nustar\ band both by attenuation of the direct beam and by recoil and photoabsorption of the scattered component. The Monte-Carlo-noise-moderated suppression floor for {\tt BNsphere} in Fig.\,\ref{fig:xmodels} should thus be regarded as conservative. In short, under the enhanced obscuration scenario, the {\em NuSTAR} non-detection requires not only depletion of the direct transmitted continuum by deeply Compton-thick columns, but also strong absorption of all scattered X-rays. Only obscurers approaching full covering can therefore provide the $\gtrsim$\,4\,dex suppression required to hide a source at its typical obscured-phase luminosity, and we will further discuss the viability of such a scenario in the following section.

Turning next to the radio, \citet{motta27} have published long-term light curves of the source demonstrating an approximately monotonic radio decline since early 2025, favouring a fading central engine. They also recover \grs\ in deep MeerKAT imaging at its faintest radio levels in 2026 July with an optically thin spectral index $\alpha=-0.67\pm0.41$; $F_\nu$\,$\propto$\,$\nu^\alpha$. Free--free absorption is the dominant opacity mechanism for an ionised cocoon in the high gas-density ($n_e$) regime. Using the standard radio approximation \citep[e.g.][]{MezgerHenderson1967,condonransom16}, 
\begin{equation}
\tau_{\rm ff} \simeq
3.28\times10^{-7}
\left(\frac{T_{\rm e}}{10^4\,{\rm K}}\right)^{-1.35}
\left(\frac{\nu}{\rm GHz}\right)^{-2.1}
\left(\frac{\rm EM}{\rm pc\,cm^{-6}}\right),
\label{eq:tauff}
\end{equation}
where the emission measure, ${\rm EM}=\int n_{\rm e}^2 dl$. For an ionised screen with $N_{\rm H}=10^{25}\,{\rm cm^{-2}}$ and thickness $R$, $n_{\rm e}$\,$\simeq$\,10$^{13}$\,$\left(\frac{N_{\rm H}}{10^{25}\,{\rm cm^{-2}}}\right)\,\left(\frac{10^{12}\,{\rm cm}}{R}\right)$\,cm$^{-3}$ and ${\rm EM}$\,$\simeq$\,$n_e^2R$, yielding a free--free optical depth of 
\begin{equation}
\tau_{\rm ff}(1.28\,{\rm GHz}) \simeq 6.3\times10^{12}\,\left(\frac{n_e}{10^{13}\,{\rm cm^{-3}}}\right)^2\left(\frac{R}{10^{12}\,{\rm cm}}\right).
\end{equation}
Thus, if even moderately ionised, such a compact Compton-thick obscurer would be overwhelmingly optically thick at 1.28\,GHz. A temperature of $T_e$\,=\,10$^4$\,K is adopted here aligned with the nebular gas properties reported in \citet{Gandhi2025}; however, the inference of a very high $\tau_{\rm ff}$ is robust to any plausible temperature $T_e$ and also across a wide range of ionisation fractions $x$. The plasma cut-off frequency is $\nu_{\rm p}$\,$\approx$\,9\,$(n_e/10^{12}\,{\rm cm^{-3}})^{1/2}$\,GHz assuming $x$\,=\,1 \citep{rybickilightman}, so any electron density above $n_{\rm e}$\,$\approx$\,2\,$\times$\,10$^{10}$\,cm$^{-3}$ would prevent the 1.28\,GHz core being visible.

The same inference also holds across a reasonable range of cocoon radii $R$. Here, it is also noteworthy that VLBI observations have typically resolved synchrotron radiation from the source on angular scales of $\mathcal{O}{\rm (mas)}$. For instance, \citet{dhawan00} found the nucleus to be resolved as a self-absorbed compact jet of length 10\,$\lambda_{\rm cm}$\,$\left(\frac{d}{12\,\rm kpc}\right)$\,AU. At $d$\,=\,9.4\,kpc, the corresponding length at $\nu$\,=\,15\,GHz ($\lambda_{\rm cm}$\,=\,2) is $\approx$\,2\,$\times$\,10$^{14}$\,cm. So if a compact jet were still being generated in \grs, any obscuring cocoon would need to have a radius $R$\,$\gtrsim$\,10$^{14}$\,cm to hide it at 15\,GHz. In the standard model of compact-jet synchrotron radiation \citep{blandford79}, the characteristic emission frequency from any given location scales inversely with distance from the jet base, so $R$ would need to be larger still in order to hide the 1.28\,GHz emission. Such large size scales place severe constraints on the viability of the cocoon, as we discuss in the following section.

Another corollary of these arguments is that the GHz-emission region of \citet{motta27} cannot be enclosed by a compact Compton-thick ionised cocoon. Eq.\,\ref{eq:tauff} implies that any such cocoon would be free--free thick by many orders of magnitude. This is consistent with the interpretation of \citet{motta27} and their economical reading that the observed optically thin spectral slope found in July 2026 results from optically thin material left-over from past activity, with MeerKAT unable to resolve legacy ejecta from any weaker underlying self-absorbed core. A similar caveat could apply to their December 2025 $S$-band core detection limited by the MeerKAT beam. No spectral index is available for this epoch, however, so its flux density is best regarded as an upper limit on any steady compact core, which we discuss below.

\subsection{Intrinsic quenching}

\subsubsection{The source cannot still be accreting at historical levels}
\label{sec:mustbequenced}

The estimates in the previous section imply that a deeply Compton-thick, ionised obscuring cocoon can easily be opaque to transmitted radiation in X-rays and also to compact-jet radio emission. Erasing all {\em scattered} X-ray emission places additional stringent constraints on the effective geometry and porosity of the cocoon along all sightlines. 

Let us assume for a moment that, albeit stringent, the above conditions are satisfied: i.e., the source is intrinsically active at/near historical levels but newly embedded within a 4$\pi$-covering Compton-thick cocoon. In such a scenario, what happens to the power ($L_{\rm abs}$) that does not escape the cocoon and is absorbed? This energy cannot simply disappear; once reprocessed, it has to emerge somewhere else. Thermal re-emission is characterised by an effective blackbody temperature 
\begin{equation}
    T_{\rm eff}\approx\left(\frac{L_{\rm abs}}{4\pi R^2\sigma_{\rm SB}}\right)^{1/4}
    \label{eq:SB}
\end{equation}
with $\sigma_{\rm SB}$ the Stefan--Boltzmann constant. $T_{\rm eff}$ is also independent of absorber geometry, because a small covering-factor absorber (a disc or torus, say) would intercept only a  fraction $\Omega$/4$\pi$ of the (isotropic) source luminosity, but then also reradiate from a correspondingly smaller area.

As an example of the constraints imposed by reprocessing, consider a cocoon with a radius $R$\,$\approx$\,10$^{13}$\,cm and illuminated at a typical historical luminosity level likely appropriate for the X-ray obscured phase, $L_{\rm hist}$\,$\sim$\,$L_{\rm abs}$\,$\approx$\,0.05\,L$_{\rm Edd}$\,(say)\,=\,7\,$\times$\,10$^{37}$\,erg\,s$^{-1}$, Eq.\,\ref{eq:SB} yields a temperature $T_{\rm eff}$\,=\,5,600\,K. The corresponding blackbody flux density from this cocoon at 5\,\micron\ can then be computed as
\begin{equation}
F_\nu \;=\; \frac{\Omega}{4\pi}\,\pi B_\nu(T_{\rm eff})\left(\frac{R}{d}\right)^{2}10^{-0.4A_\nu}.
\label{eq:Bnu}
\end{equation}
Here $\pi B_\nu(T_{\rm eff})$ is the monochromatic flux density emerging from the surface of a blackbody of temperature $T_{\rm eff}$, and $A_\nu$ the interstellar extinction at frequency
$\nu$ from the tabulated \citet{Gandhi2025} reddening law. At a frequency corresponding to 5\,\micron, say, Eq.\,\ref{eq:Bnu} yields $F_{\rm 5\ \mu m}$\,$\approx$\,175\,mJy or $F_{\rm 5\ \mu m}^{\rm reddened}$\,$\approx$\,70\,mJy, exceeding the \spx\ detection level by almost two orders of magnitude, irrespective of any caveats related to blending. A somewhat larger cocoon makes the conflict with respect to the \spx\ detection level {\em worse}; e.g. with $R$\,=\,10$^{15}$\,cm, $T_{\rm eff}$ becomes 600\,K, which pushes the peak of the reprocessed emission to lie within the thermal infrared regime of \spx, with a predicted $F_{\rm 5\ \mu m}$\,$\approx$\,10\,Jy. 

Cocoons that are even larger still could overcome this issue, but would then start to be mass-budget limited. A Compton-thick shell with radius $R$ would have a total mass of
\begin{eqnarray}
    M_{\rm shell} &=& 4\pi\ R^2\ N_{\rm H}\ \mu_{\rm H}\ m_{\rm p}\nonumber\\
    &=&14.7\ {\rm M_\odot}\ \left(\frac{N_{\rm H}}{10^{25}\ {\rm cm}^{-2}}\right)\left(\frac{R}{\rm 10^{16}\ cm}\right)^2
    \label{eq:mshell}
\end{eqnarray}
with $\mu_{\rm H}$\,=\,1.4 for typical cosmic abundance. This is of the order of the mass of the entire binary system, including the compact object \citep{Steeghs2013, Reid2023}, and is implausibly high. In fact, for any $R$\,$>$\,10$^{13.4}$\,cm, the shell mass already exceeds a conservative upper bound on the total mass of the accretion disc, $M_{\rm disc}$\,$\approx$\,10$^{-4}$\,M$_{\odot}$, which also matches the likely maximum mass expelled during prolific outflows inferred in 2023--24; see \S\,\ref{sec:sense}. Turning down \nh\ to compensate ends up making the shell Compton-thin and violating the X-ray non-detection constraint.

\begin{figure*}
    \centering
    \includegraphics[width=0.75\linewidth]{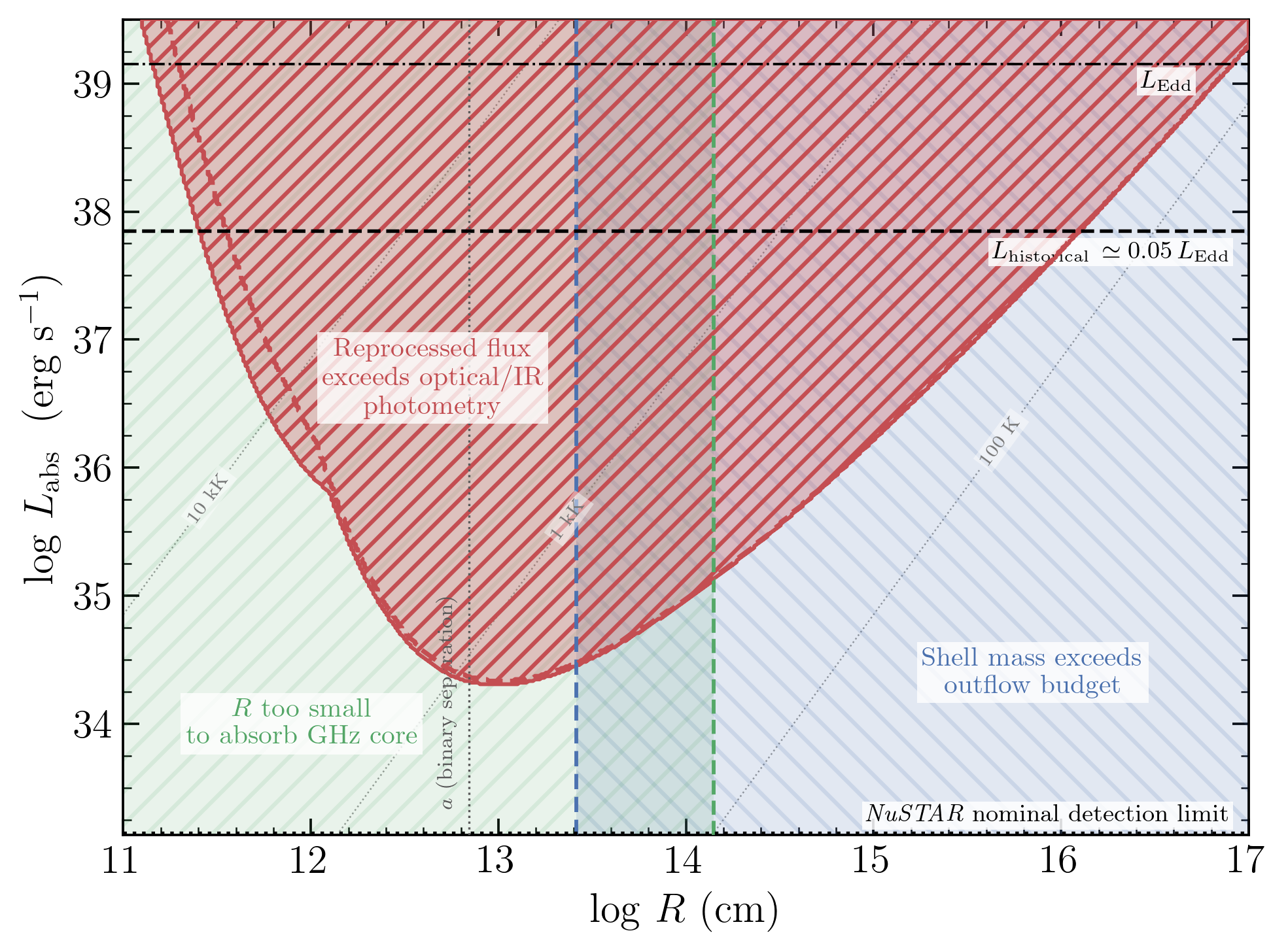}
    \caption{
    Parameter space of absorbed luminosity $L_{\rm abs}$ and radius $R$ for a Compton-thick spherical cocoon embedding a source with unabated luminous intrinsic accretion activity. Three complementary constraints are shown as hatched excluded regions: (1) The red region denotes models whose reddened thermal reprocessed emission exceeds at least one optical, near-IR or \spx\ constraint; its boundary is set primarily by the optical/near-IR data at $\log(R/{\rm cm})\lesssim12.2$ and by \spx\ at larger radii. The filled red region adopts the nominal interstellar extinction, $A_{\rm V}=19.6$ mag, while the dashed red contour shows the conservative $+1\sigma$ case, $A_{\rm V}=21.3$ mag, in order to gauge sensitivity to $A_{\rm V}$. (2) At $\log(R/{\rm cm})\gtrsim13.4$, the mass required to sustain the obscuring column exceeds the adopted outflow-mass budget of the disc ($M_{\rm disc}$\,$\approx$\,10$^{-4}$\,M$_\odot$), excluding the region on the right in blue. (3) Finally, the green hatched region on the left is that excluded based on the requirement to obscure the GHz compact-jet core known in this system (e.g., \citealt{dhawan00}). A handful of benchmark luminosities, radii and thermal temperature loci are indicated as lines to guide the eye.
    }
    \label{fig:exclusion}
\end{figure*}

These complementary multiwavelength constraints, when considered jointly, provide the most stringent boundary conditions on the allowed cocoon properties. These are shown in terms of the parameter space in $R$ and $L_{\rm abs}$ in Fig.\,\ref{fig:exclusion}. The plot considers the radii span of log$(R/{\rm cm})$\,=\,11--17, straddling plausible thermal-wind launching radii at the small end \citep{miller2025} and the maximum distance a wind travelling at $\upsilon_{\rm wind}$\,=\,10$^3$\,km\,s$^{-1}$ could have travelled since 2018 at the upper end. Note that the binary separation $a$ is $\approx$\,7\,$\times$\,10$^{12}$\,cm. Three separate constraints are implemented in the plot: (1) the lack of reprocessed emission in excess of the companion star excludes regions in the upper part of the plot, as $R$ and $T_{\rm eff}$ of the putative cocoon move the peak of a Planck curve through the MIR (\spx), the near-IR (GTC), and the optical (COLIBRI) bandpasses (with Eq.\,\ref{eq:Bnu} delivering the flux predictions); (2) regions on the right-hand side of the plot can be excluded based on mass constraints at large radii, following Eq.\,\ref{eq:mshell} above for a limiting mass budget of $M_{\rm disc}$ and \nh\,=\,10$^{25}$\,cm$^{-2}$ (conservative, since $\gtrsim$\,2\,$\times$\,10$^{25}$\,cm$^{-2}$ is needed to hide a luminous source; \S\,\ref{sec:discussion_cocoon}); and finally, (3) the left-hand side of the plot excludes radii that are too small to cocoon the GHz-compact-jet radio core expected on angular scales of (at least) 1\,mas, corresponding to 1.4\,$\times$\,10$^{14}$ cm. The reprocessing medium is optically thick to its own emission throughout Fig.\,\ref{fig:exclusion}, but the opacity source changes with $R$. For compact configurations, the cocoon is free--free thick well into the infrared [$\tau_{\rm ff}(5\,\mu{\rm m})$\,$\gtrsim$\,100 according to Eq.\,\ref{eq:tauff}], whilst at larger radii, $T_{\rm eff}$ falls below the dust-sublimation temperature so grains --- if able to condense --- would serve the same role.

When considered together, these complementary constraints end up excluding the entire parameter space of the plot. Stability arguments, not considered here, could provide further constraining leverage. A quasi-spherical cocoon is unlikely to remain both static and Compton-thick for long without continuous confinement or replenishment: compact configurations evolve on short dynamical timescales, whereas extended configurations disperse and rapidly hit the mass-budget limit discussed above.

The {\em only} plausible self-consistent solution is then to require a substantial intrinsic weakening of the accretion power itself. The source can no longer be accreting near historical levels.

\begin{figure*}
    \centering
    \includegraphics[width=0.75\linewidth]{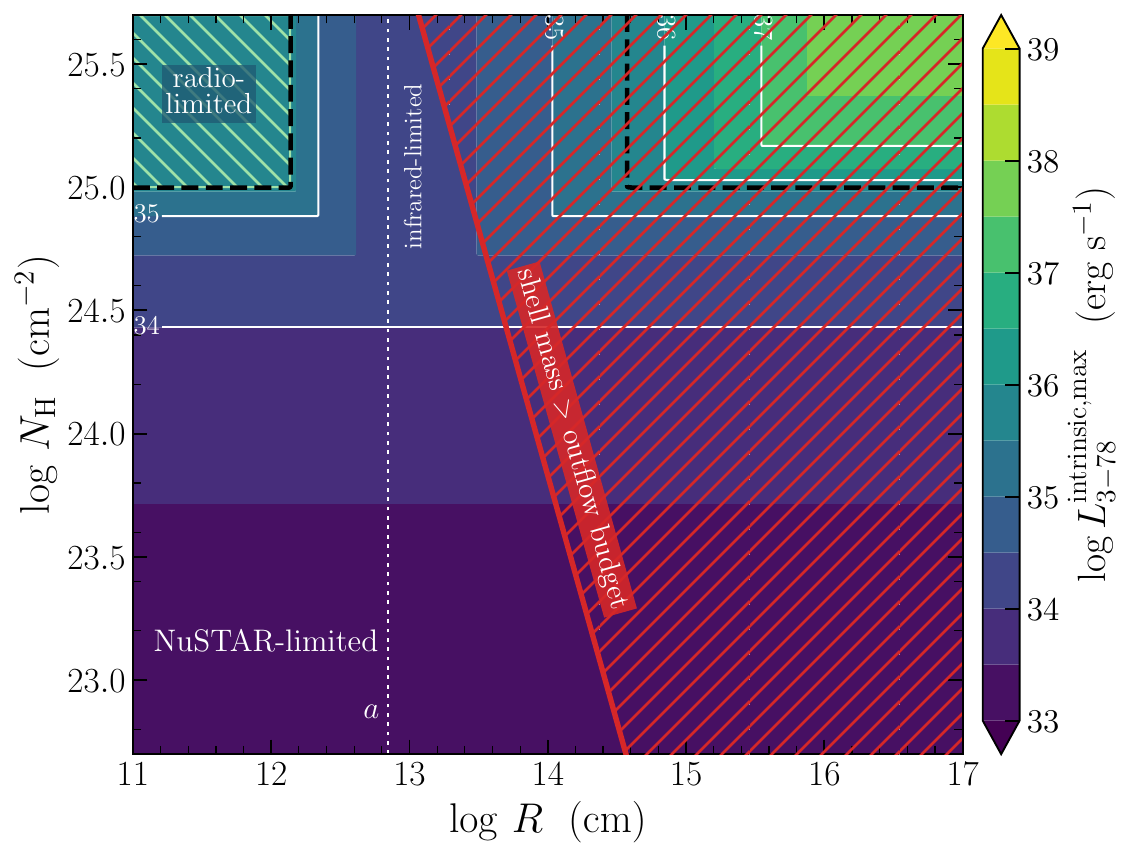}
    \caption{
    Allowed parameter space for a combination of intrinsic weakening and obscuration. This plot is similar in essence to Fig.\,\ref{fig:exclusion}, but now in terms of the joint constraints on the maximum intrinsic power \lintnustar\ (colour scale; Eq.\,\ref{eq:lmax}) that any given combination of (\nh, $R$) can sustain without violating either the \nustar\ non-detection or the \spx\ constraints. These constraints are grouped in 0.5-dex-wide zones in $\log$\,\lintnustar\ (a few key levels are annotated in white). The L-shaped zones near the top result from the \spx\ infrared constraints improving upon those provided by \nustar, especially for radii straddling $R$\,$\sim$\,10$^{13}$\,cm. The figure assumes a full-covering ($\Omega/4\pi$\,=\,1) {\tt BNsphere} obscurer, with \lintnustar\ taken as a proxy for the absorbed power $L_{\rm abs}$. The black dashed contour corresponds to \lintnustar\,=\,4.4\,$\times$\,10$^{35}$\,erg\,s$^{-1}$, the intrinsic luminosity obtained using the deep MeerKAT December 2025 detection as an anchor for the radio--X-ray correlation \citep{motta27}. The red hatched region is where the mass budget begins to be violated, and the green-hatched (radio-limited) region excludes luminous engines obscured by a cocoon that is too compact to hide the GHz emission region expected from a self-absorbed compact jet, which would then have exceeded the observed MeerKAT flux density.
    \label{fig:joint_weak_obscured_constraints}}
\end{figure*}

\subsubsection{Allowed parameter space of weaker intrinsic powers with continued obscuration}
\label{sec:degenerate}

In the absence of any residual obscuration, an X-ray weakening of the source by at least 4\,dex relative to $L_{\rm hist}$ would be required to satisfy the \nustar\ limit and infrared constraints (Fig.\,\ref{fig:lmirlx}). Relative to the Eddington luminosity, the corresponding X-ray quenching factor is $L_{\rm Edd}$/\llimnustar\,$\gtrsim$\,10$^{6}$.

A scenario with {\em joint} partial quenching and continued absorption that hides the central engine is not ruled out by the previous arguments. This is because an inherently weaker source can satisfy the observational detection limits with fewer demands on the veiling column, and with correspondingly less reprocessed power, thus rendering the intrinsic power and column degenerate. However, this degeneracy can be broken, at least in part, by again leveraging the multiwavelength constraints imposed in the previous section.

This is demonstrated in Fig.\,\ref{fig:joint_weak_obscured_constraints}, which plots the obscuring column \nh\ against the cocoon radius $R$, together with a colour scale denoting the maximum allowed intrinsic power \lintnustar\ as a third variable. Adopting the {\tt BNsphere} geometry, and taking \lintnustar\ as a proxy for $L_{\rm abs}$ (i.e., a bolometric correction of unity, which is conservative for Compton-thick columns), this maximum is
\begin{equation}
L_{\rm 3-78,\,max}^{\rm int}(N_{\rm H},R)\simeq\min\left[\mathcal{S}(N_{\rm H})\,L_{3-78}^{\rm obs,lim},\ L_{\rm abs,\,max}(R)\right],
\label{eq:lmax}
\end{equation}
where \supp\ is the 3--78\,keV flux suppression of Fig.\,\ref{fig:xmodels}a, \llimnustar\ is the nominal \nustar\ limit (\S\,\ref{sec:nustar}), and $L_{\rm abs,\,max}(R)$ is the largest absorbed power whose reprocessed emission (Eqs.\,\ref{eq:SB} and \ref{eq:Bnu}) does not exceed the \spx\ constraints. When the intrinsic luminosity is reduced (darker colours), so is the column needed to hide it while satisfying the \nustar\ detection limit. Larger columns (towards the top) are effective in hiding much larger intrinsic luminosities (lighter colours). Complementary to this are the constraints on the {\em reprocessed} powers provided by \spx, which is optimally sensitive around $T_{\rm eff}$\,$\sim$\,1,000\,K, corresponding to radii $R$\,$\approx$\,10$^{13}$\,cm for $L_{\rm abs}$\,$\sim$\,10$^{35}$\,erg\,s$^{-1}$ (Eq.\,\ref{eq:SB}; cf.\ Fig.\,\ref{fig:exclusion}). This extra constraining power explains the wedge of lower allowed \lintnustar\ values jutting upwards around $R$\,$\approx$\,10$^{13}$\,cm. The mass constraint from Fig.\,\ref{fig:exclusion} is also plotted here\footnote{At fixed $M_{\rm shell}$\,=\,$M_{\rm disc}$, \nh\ scales as $R^{-2}$ according to Eq.\,\ref{eq:mshell}, so the boundary is sloped instead of vertical.}, and excludes the larger radii zones.

A weakened and obscured source scenario has also been proposed by \citet{motta27}, with their December 2025 radio detection used to anchor the source on the radio--X-ray correlation of \citet{crook-mansour2026} and infer a residual column \nh\,$\gtrsim$\,7\,$\times$\,10$^{24}$\,cm$^{-2}$ hiding ongoing low-level accretion at a level of \lintnustar\,$\sim$\,4.4\,$\times$\,10$^{35}$\,erg\,s$^{-1}$. This interpretation is contingent on the association of their radio detection with current accretion activity, as opposed to past optically thin ejecta. Furthermore, their column and inferred intrinsic power are based on attenuation of the directly transmitted photons only. In our {\tt BNsphere} geometry, photons Compton-scattered into the line-of-sight also contribute to the observed flux, so the requisite suppression (\supp\,$\approx$\,300) is only reached at \nh\,$\approx$\,10$^{25}$\,cm$^{-2}$, and the corresponding \lintnustar\ is shown as the dashed black contour. Notably, this is not a horizontal line independent of $R$, because \spx\ imposes meaningful restrictions limiting the maximum allowed obscurer radii even at these relatively low luminosities.

The green-hatched (radio-limited) region marks where the radio constraint is binding: a cocoon smaller than $\approx$\,1.4\,$\times$\,10$^{14}$\,cm (1\,mas, or $\approx$\,9\,AU) cannot hide the GHz emission region of a compact jet. This is analogous to constraint (3) in Fig.\,\ref{fig:exclusion}, but now applied only to powers that would be radio-detectable. Treating the December 2025 MeerKAT flux density as an upper limit on any compact core, a source on the radio--X-ray correlation would exceed it for \lintnustar\,$>$\,4.4\,$\times$\,10$^{35}$\,erg\,s$^{-1}$, so such powers are excluded for compact cocoons, and \lintnustar\ is capped at this value.

The key result here is the set of allowed combinations of \nh, $R$ and \lintnustar\ in Fig.\,\ref{fig:joint_weak_obscured_constraints} (outside the hatched regions). Powers of up to \lintnustar\,$\approx$\,4\,$\times$\,10$^{35}$\,erg\,s$^{-1}$ (still more than two dex below $L_{\rm hist}$) are not excluded, if obscured by a compact Compton-thick column. The corresponding minimal X-ray quenching factor in this case is $L_{\rm Edd}$/\lintnustar\,$\approx$\,3\,$\times$\,10$^{3}$ or $\approx$\,2\,dex below $L_{\rm hist}$. More extended obscurers are possible with weaker intrinsic accretion powers; and at even lower intrinsic levels approaching the \nustar\ detection limit, Compton-thin columns would also suffice.

Fig.\,\ref{fig:joint_weak_obscured_constraints} assumes a fixed $\Omega/4\pi$\,=\,1, and thus does not consider the influence of a smaller covering factor. A non-4$\pi$ obscurer, such as a partially covered sphere, a tilted ring or warped outer disc, would reprocess a smaller fraction of emitted radiation than a spherical cocoon, reducing the predicted infrared flux in proportion to $\Omega/4\pi$ (Eq.\,\ref{eq:Bnu}). Conversely, such obscurers also place more demands on the geometrical configuration relative to the observer, in order to ensure that the line-of-sight to the nucleus remains hidden. They also leave illuminated surfaces in view, so that \supp\ saturates at $\lesssim$\,3.8\,dex for the partial-covering geometries tested in \S\,\ref{sec:discussion_cocoon}, limiting any hidden power to \lintnustar\,$\lesssim$\,10$^{37}$\,erg\,s$^{-1}$; but a compact warp cannot hide GHz jet emission arising on scales $\gtrsim$\,10$^{14}$\,cm, so the radio cap above applies irrespective of $\Omega$. Fig.\,\ref{fig:joint_weak_obscured_constraints} shows that we cannot rule out small-scale obscurers from the constraints considered thus far. Compton-thick disc warps could begin around $R$\,$\approx$\,10$^{11}$\,cm and stretch beyond $R$\,$\approx$\,10$^{12}$\,cm \citep[e.g.,][]{miller2025}. Given the jet inclination angle of 64$^\circ$ \citep{Reid2023}, a non-negligible covering factor $\Omega/4\pi\gtrsim0.1$ (say) would likely be required for any vertically extended or strongly warped disc to intercept the line-of-sight, and is also consistent with the density-dependent solid angles inferred for the photoionised line-emitting gas seen by {\em XRISM} \citep{miller2025}. Detailed modelling of warped disc geometries will be necessary, and we also discuss follow-on observational tests of various scenarios in \S\,\ref{sec:future}.

\section{Making sense of it all}
\label{sec:sense}

Fig.\,\ref{fig:timeline} shows a schematic depiction of the life cycle of \grs, from discovery in 1992, through the `X-ray obscured' phase starting $\approx$\,2018, transitioning around 2025 to the current low-luminosity state. The detailed evolution of the source since 2018 remains far from clear. Some unknown perturbation clearly caused it to enter the `X-ray obscured' phase, and it has continued to decline in {\em apparent} X-ray emission since, albeit sporadically punctuated by impressive multiwavelength activity. The deep detections and limits analysed herein now suggest a likely progression towards a new state with {\em intrinsically} weak or absent accretion activity and resembling quiescence. While  definitive conclusions  on the return to quiescence can only be verified based on source behaviour in the long term, there are multiple strands of evidence already supporting this, including (1) the fact that the current infrared emission is entirely consistent with the companion star alone, and has remained consistent at this level since September 2025; (2) the location of the source in the \lmir--\lx\ plane (Fig.\,\ref{fig:lmirlx}), where it is seen to occupy the regime of other quiescent black-hole X-ray binaries; (3) consistency of the current multiwavelength data with orbital-period scalings when compared with other quiescent sources \citep{motta27}; (4) the substantive monotonic decline in the radio and X-rays since early 2025 \citep{motta27}; and (5) the fact that substantial fractions of parameter space for obscured scenarios are beginning to be excluded (Figs.\,\ref{fig:exclusion}, \ref{fig:joint_weak_obscured_constraints}).

During the prolific flaring of 2023--24, \citet{Gandhi2025} suggested the presence of a strong transient outflow based on modelling of the \jwst\ MIRI continuum, with inferred mass-loss rates of $\dot{M}_{\rm wind}$\,$\sim$\,10$^{-4}$\,M$_\odot$\,yr$^{-1}$ ---  powerful enough to expel the accretion disc on a timescale of $\sim$\,months (assuming a disc mass of $M_{\rm disc}$\,$\approx\,5\,\times\,10^{-6}-5\,\times\,10^{-5}\,M_{\odot}$; \citealt[][]{truss06, koljonen21}). This was supported by the detection of extremely fast optical wind signatures in the form of He\,I P\,Cygni absorption signifying outflow speeds $\upsilon_{\rm out}$\,$\sim$\,10$^3$\,km\,s$^{-1}$, and even broader hydrogen recombination lines, preceding the \jwst\ observation epoch by just one week \citep{sanchezsierras23}.

The late-stage decline was also far from monotonic, with substantial radio activity as late as 2025. RATAN-600 multi-frequency monitoring recorded a sequence of major radio
flares during 2025 January--February \citep{trushkin25}, with contemporaneous
Tianma~65\,m and Sheshan~25\,m monitoring measuring very bright peak flux densities of
$\approx$\,1.5\,Jy at both 2.25 and 8.42\,GHz around MJD\,60688--60691
\citep{li2026}. EAVN imaging resolved jet structure on mas scales during
the flares \citep{yan2026_distinctjet}. A $\approx$\,0.03\,Hz radio quasi-periodic
oscillation, together with its $0.06$\,Hz harmonic, was subsequently detected on MJD~60765 and 60772 (2025 March 31 and
April 7; \citealt{li2026}). A structured, jet-producing accretion flow was
therefore present only $\approx$\,7--10 months before our {\em NuSTAR} non-detection. These various quoted durations of time are all much longer than the rapid outburst climb and quenching of V404\,Cyg, which took all of two weeks to complete this cycle in 2015 \citep{rodriguez15, Walton2017, Gandhi2016}. But \grs\ also hosts a much larger accretion disc reservoir than V404\,Cyg. 

We propose that outflows during the flaring activity across the period of $\approx$\,2023--2025 resulted in transient but explosive clear-out of the inner accreting fuel supply, at a strongly variable rate $\dot{M}_{\rm wind}$, resulting in the current apparent quenching of accretion activity. Though we are not aware of any reported direct detections of outflows during the giant flaring of 2025 to corroborate and test this hypothesis, the source was observed with the microcalorimeter mission \xrism\ on MJD 60600 (2024 October 17) in the lead-up to the radio-flaring peaks, and \citet{miller2025} reported the presence of a continued (albeit slow) X-ray wind at that time.

The current intrinsic luminosity of the source remains uncertain, being partly degenerate with obscured-source scenarios (\S\,\ref{sec:degenerate}), and we will discuss future observational tests to break some of these degeneracies in \S\,\ref{sec:future}. If the deep radio detection of \citet{motta27} from December 2025 reflects contemporaneous nuclear activity, the accretion rate necessary to sustain the implied intrinsic power can be estimated as $\dot M_{\rm acc}$\,$\approx$\,\lintnustar/$\eta c^{2}$ (assuming a \nustar\ bolometric correction factor of $\mathcal{O}(1)$). For a radiative efficiency $\eta$\,=\,0.1, this is only $\dot M_{\rm acc}$\,$\sim$\,8\,$\times$\,10$^{-11}$\,\Msun\,yr$^{-1}$\,$\equiv$\,3\,$\times$\,10$^{-4}$\,$\dot M_{\rm Edd}$. Alternatively, at these very low accretion rates, the decline could reflect a more conventional end-of-outburst transition to a radiatively inefficient accretion flow (RIAF) in which \lintx\,$\propto$\,$\dot{M}^{2}$, so that a modest fall in accretion rate produces a steeper X-ray decline \citep{narayan95,esin97}. Nevertheless, our central conclusion --- that the current decline must be intrinsic --- holds in either case.

\subsection{The outburst life cycle of GRS 1915+105}
\label{sec:life}

The interval spanning source discovery in 1992 through to the \nustar\ non-detection epoch  of November 2025 implies an outburst duration of $\lesssim$\,33.3\,years. Alternatively, the fact that the first \spx\ epoch in September 2025 shows no obvious excess above the companion-star spectrum suggests that the transition to the current quenched state may have been completed earlier. Assuming that an active accretion flow was present at the time of the giant RATAN-600 radio flaring in early 2025 implies that the final transition took less than a year. An upper limit to the accreted mass across the outburst can be estimated from the mean accretion rate $\dot M_{\rm acc}=\langle L\rangle/\eta c^{2}$. Taking $\langle L\rangle$\,$\sim$\,0.5\,L$_{\rm Edd}$ as a representative approximation with $\eta$\,=\,0.1, we obtain $\dot M_{\rm acc}$\,$\sim$\,1.3\,$\times$\,10$^{-7}$\,M$_{\odot}$\,yr$^{-1}$, which aggregates to a total of $M_{\rm acc}$\,$\sim$\,4\,$\times$\,10$^{-6}$\,M$_{\odot}$ within 33.3\,yr. This accumulated mass could easily have been expelled by the very strong outflows in 2023 reported by \citet{Gandhi2025}, even if those outflows operated for just a fraction of a year. A maximally spinning black hole with larger $\eta$ only strengthens this inference.

The eventual end of the outburst in \grs\ as a result of 
exhaustion of the finite disc mass was also anticipated by \citet[][see also \citealt{deegan2009}]{truss06}, who showed that the duration of the outburst is a sensitive probe of the radius $R_{\rm irr}$ within which disc irradiation dominates hydrogen ionisation. A total outburst duration of $t_{\rm outburst}$\,$\approx$\,33\,yr is plausibly consistent with $R_{\rm irr}$\,$\approx$\,1--2\,$\times$\,10$^{12}$\,cm, for a range of outflow rates $\dot{M}_{\rm wind}$ up to the Eddington accretion rate (see their Table\,1). This estimate should now be revised to account for newer estimates of the system physical parameters (e.g., the newer $q$ by \citeauthor{Steeghs2013} is 60\% of that adopted by \citeauthor{truss06}, which in turn implies a larger circularisation radius $R_{\rm circ}$\,$\approx$\,3\,$\times$\,10$^{12}$\,cm and thus a larger disc reservoir; cf. Eq.\,4.20 of \citealt{frankkingraine}). Additional dependences that need to be accounted for include the disc irradiation efficiency (further complicated by changing obscuration) and the wind mass-loss rate (which the \jwst\ observations showed to be transiently higher than the Eddington rate by orders of magnitude). Clearly, a full exploration of parameter space is warranted to constrain  disc depletion models.

The root trigger of the late-stage changes remains unclear. A rapid change in the inner accretion-flow orientation has been proposed, with radio imaging suggesting that the jet/disc axis of \grs\ changed substantially during the obscured state and may have placed the disc close to the line-of-sight near the {\em JWST} epoch \citep{jiang2026, Rodriguez2025}. More generally, warped or precessing inner discs have also been invoked to explain recent jet-orientation changes in the source \citep{yan26, deregt2026}. The possibility of stellar interaction within a hierarchical triple system has also been raised \citep{Rodriguez2025}. It will be critical to continue monitoring the source to assess how long this phase lasts, and whether any other unexpected perturbations follow.

Once depleted, the time for activity to resume is set by two scales. The giant donor transfers mass at $|\dot M_2|\approx4\times10^{-9}\,M_\odot\,{\rm yr^{-1}}$ (cf., Eq.\,8 of \citealt{King1997}), or up to $\sim$\,2\,$\times$\,10$^{-8}$\,M$_\odot$\,yr$^{-1}$ \citep{truss06, deegan2009}. The Roche-lobe refilling timescale (i.e., for mass transfer to replenish the disc lost from within the black hole's Roche lobe) is 
\begin{equation}
t_{\rm RL} = \frac{M_{\rm lost}}{|\dot M_2|} \approx 10^3\ {\rm yr} \left(\frac{M_{\rm acc}}{4\times 10^{-6}\,M_\odot}\right) \left(\frac{|\dot M_2|}{4\times10^{-9}\,M_\odot\,{\rm yr^{-1}}}\right)^{-1} 
\label{eq:trl}
\end{equation}
assuming that the accreted disc mass accounts entirely for the mass loss. This rises to $t_{\rm RL}$\,$\lesssim$\,2.5\,$\times$\,10$^{4}$\,yr if the wind managed to expel a much larger $M_{\rm lost}$\,$\sim$\,$M_{\rm disc}$\,$\lesssim$\,10$^{-4}$\,M$_\odot$ during 2023--24.

\begin{figure*}
    \centering
    \includegraphics[width=1.0\textwidth]{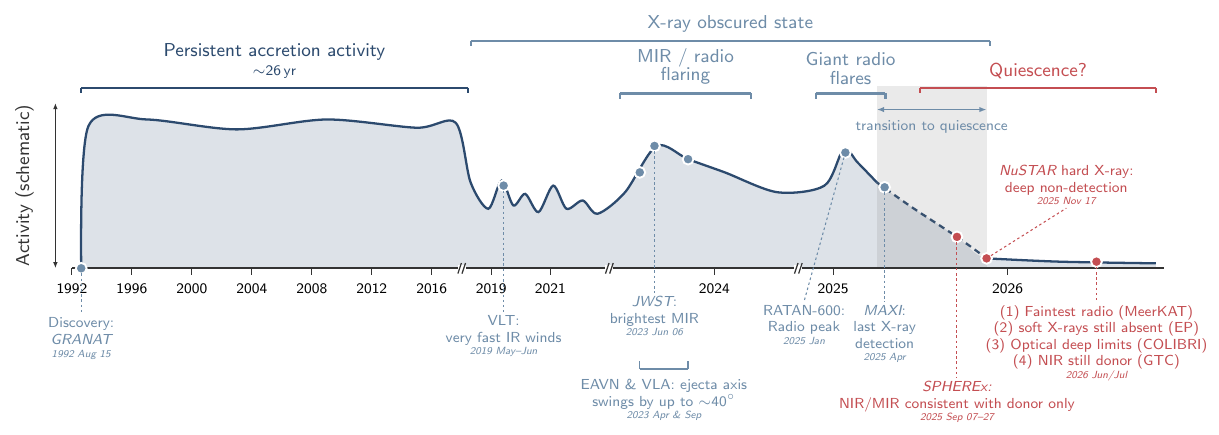}
    \caption{Schematic life history of the one (and only) outburst of \grs\ so far. The source remained persistently active with accretion rates close to Eddington for about 26 years since discovery in 1992. This was followed by a decline in the apparent X-ray flux and characterised by a rise in absorption. This `X-ray obscured phase' included dramatic periods of multiwavelength flaring, especially in the radio and the mid-infrared, and included detections of significant outflows. The transition to the current low-luminosity quenched state occurred at some point during 2025. The multiwavelength detections and limits from late 2025 onwards all sample the source in the faintest state since discovery. The y-axis should be interpreted as an approximate indicator of activity, not as a light-curve; realistic light-curve profiles will vary with wavelength (e.g., the latter phases are dominated by radio and infrared flaring and low apparent X-ray fluxes, in contrast to the earlier X-ray--bright persistent activity). Key milestone dates and phases are annotated.}
    \label{fig:timeline}
\end{figure*}

Independently, gas deposited near the circularisation radius diffuses inward through the
cold quiescent disc on the viscous timescale of 
$t_{\rm visc}\sim\alpha^{-1}(R/H)^{2}t_{\rm dyn}$,
with a dynamical timescale of $t_{\rm dyn}=\sqrt{R^{3}/GM_{\rm BH}}$, a scale height $H=c_{\rm s}t_{\rm dyn}$, and 
sound speed $c_{\rm s}=\sqrt{kT/\mu m_{\rm p}}$, and $\mu$ the mean molecular weight for cold gas, so that
{\footnotesize \begin{eqnarray}
t_{\rm visc} & \sim & \frac{\mu m_{\rm p}}{\alpha kT}\sqrt{GM_{\rm BH}R}\nonumber\\
&\approx&560\ {\rm yr}
\left(\frac{0.02}{\alpha}\right)
\left(\frac{3000\,{\rm K}}{T}\right)
\left(\frac{\mu}{1.3}\right)
\left(\frac{M_{\rm BH}}{11.2\,M_\odot}\right)^{1/2}
\left(\frac{R}{3\times10^{12}\,{\rm cm}}\right)^{1/2}.
\label{eq:tvisc}
\end{eqnarray}}
Irradiation is negligible at the present luminosity and does not shorten these estimates. Disc-instability simulations of this system likewise predict recurrence times of $\sim10^{4}$\,yr \citep{deegan2009}. Renewed activity is, therefore, not expected for decades to centuries (or longer), unless the outer disc has been left close to its critical surface density. It is to be noted, however, that rebrightening events have been seen in several systems following short periods of return to quiescence \citep[e.g.,][]{zhang2019_rebrightening}, and it remains to be seen whether any such rebrightening will occur for \grs. The late peaks in the timeline of \grs\ (Fig.\,\ref{fig:timeline}), especially the one in 2025, may also represent a (large) reflare as the fuel supply for accretion approached depletion.

\subsection{Further Tests and Follow-up Opportunities}
\label{sec:future}

Our SED sampling has gaps in the far-infrared to sub-mm regime. Searching for reprocessed emission could provide a sensitive test of any residual veiling cocoon. This could include searches for cool atomic and molecular emission lines as well as continuum reprocessing signatures. Any cool extended cocoon could be resolved with ALMA with angular diameter 
\begin{equation}
    \theta=44.9\,\left(\frac{t_{\rm expansion}}{\rm yr}\right)\,\left(\frac{\upsilon_{\rm out}}{\rm 10^3\,km\,s^{-1}}\right)\,\left(\frac{\rm 9.4\,kpc}{d}\right)\,{\rm mas}, 
\end{equation}
assuming that the gas has continued to outflow at $\upsilon_{\rm out}$\,$\sim$\,10$^3$\,km\,s$^{-1}$ for time $t_{\rm expansion}$. Detectability will depend strongly on surface brightness, but resolving the angular extent itself will be within reach of ALMA for $t_{\rm expansion}$ of order a few years during the X-ray obscured phase.

Continued radio monitoring will be critical to distinguish between sustained low-level core accretion activity pointing to the birth of a RIAF on the radio/X-ray correlation \citep{Gallo2003,Corbel2013,Gallo2006} vs. expanding ejecta that should fade on a timescale of years. Should future MeerKAT monitoring find continued radio fading in the long term, that would point to an even more dramatic collapse of the accretion flow, obviating the need for Compton-thick obscuration and supporting a bare line-of-sight to the nucleus.

X-ray observations in the current state should also be able to improve upon existing limits. The joint Einstein Probe limit of $F_{1-10}^{\rm obs}$\,$<$\,8.1\,$\times$\,10$^{-14}$\,erg\,s$^{-1}$\,cm$^{-2}$ from mid-2026 \citep{motta27} is based on a 7.6\,ks stack, so modest \chandra\ or \xmm\ observations could lower it by an order of magnitude, to $L_{\rm X}$\,$\sim$\,10$^{32}$\,erg\,s$^{-1}$, which could then be combined with our joint constraints on \lintx\ and \nh\ (Fig.\,\ref{fig:joint_weak_obscured_constraints}) to better constrain any obscuration properties. A detection with only interstellar absorption would favour an unobscured, faded source, or help to constrain scattering and reflection fractions. A non-detection would require any leaked or scattered fraction of a radio-implied engine to be $\lesssim$\,10$^{-3}$, increasingly favouring continued intrinsic fading. Tests for a warped disc are more strongly model-dependent; one possibility is to search for variability in flux or reflection fraction on the warp-relaxation or orbital timescales of weeks \citep[e.g.][]{nixon2016}, although such an interpretation would be difficult to establish uniquely.

\spx\ will continue to sample the source every few months, and detection of any stochastic infrared flux or spectral  variability will be a smoking-gun signature of ongoing accretion-related activity. We also expect the mission pipeline processing to improve, which could help to better deconvolve neighbouring source contamination (something that better resolved imaging with ground-based facilities will also provide). Concurrently, infrared spectroscopy of quiescent hydrogen recombination lines (e.g. Pa$\beta$, Br$\gamma$) could constrain the extent of the outer disc \citep[e.g.,][]{casares2016}.

A decline in accretion presents an opportunity to study the donor star and binary orbital characteristics unimpeded by disc veiling. This could include a search for ellipsoidal photometric modulations to determine the inclination angle of the binary orbit to the line-of-sight \citep[cf.,][]{Cantrell2010}, independent of the radio-jet inclination. Given recently reported perturbations of the jet direction, this could help settle any doubts on the system inclination. The best published fractional error on the orbital period of 4.7\,$\times$\,10$^{-3}$ would result in an accumulated present-day phase uncertainty of $\approx$\,80\,\% (\citealt{Steeghs2013}; see also \citealt{hurley2013}), and so is worth an updated check. The mass-ratio $q$ of the binary could also be worth revisiting. For a Roche-lobe filling, synchronously rotating companion star with effective radius $R_{\rm L}$,
\begin{equation}
    \frac{\upsilon\,{\rm sin}\,i}{K_2}\,=\,(1+q)\,\frac{R_{\rm L}}{a}
\end{equation}
where $K_2$ is the radial-velocity semi-amplitude and $a$ is the semi-major axis of the binary. Using the \citet{paczynski1971} approximation to $\frac{R_{\rm L}}{a}$ results in a scaling $q$\,$\propto$\,$(\frac{\upsilon\,{\rm sin}\,i}{K_2})^3$ in the limit $q$\,$\ll$\,1. The cubic exponent scaling is the reason why the current error budget in $q$ is almost 60\%\  ($q$\,=\,0.042\,$\pm$\,0.024; \citealt{Steeghs2013}), dominated by uncertainties on the measured rotational broadening $\upsilon$\,sin\,$i$\,=\,21\,$\pm$\,4\,km\,s$^{-1}$. Improving $q$ could be possible with high signal-to-noise, high resolution, near-infrared spectroscopy targeting photospheric lines; the absence of disc contamination now should facilitate these measurements. Finally, and more ambitiously, there may be a hope to resolve the binary orbit astrometrically. The semi-major axis of the donor projects an angle on the sky of $\approx$\,48\,$\mu$as, or a peak-to-peak wobble of 96\,$\mu$as in the major axis direction. With no accretion-disc contribution, the photocentre follows the donor's orbit directly. These scales should be within grasp of planned upgrades to GRAVITY+ \citep{gravityplus, gravityplus_upgrades}. Phase coverage across the binary orbit will give direct estimates of the inclination and longitude of the ascending node, allowing the orbital angular-momentum axis to be compared against the radio jet.

\section{Conclusions}
We have presented multi-faceted evidence of dramatic intrinsic weakening of accretion activity in the prototypical microquasar \grs\ for the first time since its discovery in 1992. This includes a deep hard X-ray non-detection by \nustar, and consistency between the observed near/mid-infrared flux levels and expectations from the companion star only (\S\,\ref{sec:data}) --- a first occurrence for both instances.

Whilst a ramp-up in circumnuclear absorption can explain the lack of transmitted X-ray emission and the current absence of a self-absorbed compact jet in the radio, this scenario requires an exceedingly low scattering fraction $f_{\rm scatt}$\,$<$\,10$^{-4}$ together with a deeply Compton-thick line-of-sight column in order to also account for the complete lack of scattered X-rays (Figs.\,\ref{fig:lmirlx}, \ref{fig:xmodels}). This, in turn, is then in conflict with the lack of reprocessed thermal emission that ought to emerge in the infrared, and with the radio and mass-budget limits on the size of any such cocoon (\S\,\ref{sec:discussion_cocoon}; Fig.\,\ref{fig:exclusion}). Instead, an intrinsic source weakening by four orders of magnitude (or more) relative to its typical historical power output is far more plausible if the nucleus is now unobscured (or by at least two orders of magnitude if compact Compton-thick obscuration persists and the detected radio core reflects ongoing low-level accretion instead of optically thin ejecta), and is also self-consistent with the location of the source in the \lmir--\lx\ plane occupied by quiescent sources (Figs.\,\ref{fig:lmirlx}, \ref{fig:joint_weak_obscured_constraints}; \S\,\ref{sec:mustbequenced}, \ref{sec:degenerate}). The joint parameter space of intrinsic power, obscuring column and obscurer extent in Fig.\,\ref{fig:joint_weak_obscured_constraints} provides strong constraints for any future model to satisfy.

A schematic timeline of the history of the one (and only) outburst of \grs\ so far is presented in Fig.\,\ref{fig:timeline}. We propose that violent flaring activity and outflows witnessed in 2023 (followed by further flaring in 2024--25) expelled much of the inner accretion disc and we are now seeing the first clear quenching following $\approx$\,33\,years of accretion activity (\S\,\ref{sec:sense}). We suggest several follow-up observational possibilities to test this scenario against that of a deeply buried cocoon, and to study the companion star and binary orbit in its present state with minimal disc veiling (\S\,\ref{sec:future}). Continued monitoring should give unique insight into any rebrightening episodes and any other rapid secular changes that follow. \grs\ has surprised us before, and may yet surprise us again.

\section*{Acknowledgements}

This research has made use of data obtained through the High Energy Astrophysics Science Archive Research Center (HEASARC) Online Service, provided by the NASA/Goddard Space Flight Center. This work made use of data from the \nustar\ mission, a project led by the California Institute of Technology, managed by the Jet Propulsion Laboratory, and funded by the National Aeronautics and Space Administration. Data analysis was performed using the {\em NuSTAR} Data Analysis Software (NuSTARDAS), jointly developed by the ASI Science Data Center (SSDC, Italy) and the California Institute of Technology (USA). {\tt nH} is a web version of the {\tt nH} {\tt FTOOL}, which was developed by Lorella Angelini at the HEASARC; the web interface is maintained by Edward J. Sabol of the HEASARC. This research has made use of the VizieR catalogue access tool, CDS, Strasbourg Astronomical Observatory, France ({\tt DOI: 10.26093/cds/vizier}).

We acknowledge the developers of {\tt NumPy} \citep{numpy}, {\tt Pandas} \citep{pandas}, {\tt Matplotlib} \citep{hunter2007}, {\tt SciPy} \citep{Virtanen2020}, and {\tt Astropy} \citep{astropy}. This research has made use of NASA's {\tt WebPIMMS} mission count rate simulator, developed by Koji Mukai at the HEASARC \citep{pimms}. This work made use of {\tt OverCite} \citep{Shariat2026}, an in-editor citation tool for \LaTeX.

This work makes use of data products from the \spx\ mission, a NASA Astrophysics Explorer mission \citep{spxmission, spxpipeline}. \spx\ is managed by the Jet Propulsion Laboratory, California Institute of Technology, with science operations and data processing by the \spx\ Science Data Center at Caltech/IPAC. This research also made use of the NASA/IPAC Infrared Science Archive, which is funded by NASA and operated by Caltech.

PG acknowledges funding from The Royal Society (SRF$\_$R1$\_$241074), and thanks UKRI Science and Technology Facilities Council (STFC) for support. He acknowledges discussions with J. Miller and S. Motta on the state of the source in the months prior to first submission, and discussions with R. Lau and the \spx\ mission team on \spx\ data and analysis. J. Miller was the PI of the \nustar\ observation presented herein. PG is also grateful to T. Maccarone for helpful and rapid comments on an initial draft. We are grateful to an anonymous reviewer for a thorough and helpful report, and for suggesting the development of Fig.\,\ref{fig:exclusion} and related discussion. Figs.\,\ref{fig:joint_weak_obscured_constraints} and \ref{fig:timeline} were also created following on from adjacent reviewer comments.

\section*{Data Availability}

The primary data utilised here are available in public archives. The respective Digital Object Identifiers (DOIs) for the key datasets are: {\tt 10.26131/IRSA652} for \spx\ QR2, and {\tt 10.26131/IRSA662} for QR3.

\bibliographystyle{mnras}
\bibliography{rev} 

\bsp	
\label{lastpage}
\end{document}